\documentclass[tighten]{aastex62}

\usepackage{multirow}
\usepackage{amsmath}

\newcommand{\fermi}{\textit{Fermi}}
\newcommand{\gaia}{\textit{GAIA}}
\newcommand{\nicer}{\textit{NICER}}

\newcommand{\swift}{\textit{Swift}}

\newcommand{\chandra}{\textit{Chandra}}

\newcommand{\xmmlong}{\textit{XMM-Newton}}
\newcommand{\xmm}{\textit{XMM}}
\newcommand{\rosat}{{\em ROSAT}}
\newcommand{\nrt}{{\em Nan\c{c}ay Radio Telescope}}

\newcommand{\nh}{\mbox{$n_{\rm H}$}}

\newcommand{\chisqnu}{\mbox{$\chi^2_\nu$}}

\newcommand{\Msun}{\mbox{$M_\odot$}}

\newcommand{\simlt}{\mathrel{\hbox{\rlap{\hbox{\lower4pt\hbox{$\sim$}}}\hbox{$<$}}}}
\newcommand{\simgt}{\mathrel{\hbox{\rlap{\hbox{\lower4pt\hbox{$\sim$}}}\hbox{$>$}}}}

\newcommand{\ee}[1]{\mbox{$10^{#1}$}}
\newcommand{\tee}[1]{\mbox{$\times 10^{#1}$}}
\newcommand{\ud}[2]{\mbox{$^{+ #1}_{- #2}$}}
\newcommand{\ppm}{\mbox{$\pm$}}
\newcommand{\unit}[1]{\mbox{$\rm\,#1$}}
\def\deg{\hbox{$^\circ$}}
\def\arcdeg{\hbox{$^\circ$}}
\def\arcmin{\hbox{$^\prime$}}
\def\arcsec{\hbox{$^{\prime\prime}$}}
\def\sec{\mbox{$\,{\rm s}$}}
\newcommand{\G}{\mbox{$\,{\rm G}$}}
\newcommand{\msun}{\mbox{$\,{\rm M}_\odot$}}

\newcommand{\K}{\hbox{$\,{\rm K}$}}

\newcommand{\keV}{\mbox{$\,{\rm keV}$}}

\newcommand{\ksec}{\mbox{$\,{\rm ks}$}}
\newcommand{\msec}{\mbox{$\,{\rm ms}$}}
\newcommand{\yr}{\mbox{$\,{\rm yr}$}}
\newcommand{\kpc}{\mbox{$\,{\rm kpc}$}}
\newcommand{\pc}{\mbox{$\,{\rm pc}$}}
\newcommand{\persec}{\mbox{$\,{\rm s^{-1}}$}}

\newcommand{\percmsq}{\mbox{$\,{\rm cm^{-2}}$}}

\newcommand{\cgsflux}{\mbox{$\,{\rm erg\,\percmsq\,\persec}$}}
\newcommand{\cgslum}{\mbox{$\,{\rm erg\,\persec}$}}

\shorttitle{NICER Observations of MSPs}
\shortauthors{Guillot et al.}

\begin{document}

\title{\textit{NICER} X-RAY OBSERVATIONS OF SEVEN NEARBY ROTATION-POWERED MILLISECOND PULSARS}


\author[0000-0002-6449-106X]{Sebastien Guillot}
\affil{IRAP, CNRS, 9 avenue du Colonel Roche, BP 44346, F-31028 Toulouse Cedex 4, France}
\affil{Universit\'{e} de Toulouse, CNES, UPS-OMP, F-31028 Toulouse, France.}

\author{Matthew Kerr}
\affil{Space Science Division, Naval Research Laboratory, Washington, DC 20375, USA}

\author[0000-0002-5297-5278]{Paul S.~Ray}
\affil{Space Science Division, Naval Research Laboratory, Washington, DC 20375, USA}

\author[0000-0002-9870-2742]{Slavko Bogdanov}
\affiliation{Columbia Astrophysics Laboratory, Columbia University, 550 West 120th Street, New York, NY, 10027, USA}

\author[0000-0001-5799-9714]{Scott Ransom}
\affil{National Radio Astronomy Observatory, 520 Edgemont Road, Charlottesville, VA 22903, USA}

\author[0000-0003-1226-0793]{Julia S.~Deneva}
\affiliation{George Mason University, resident at the Naval Research Laboratory, Washington, DC 20375, USA}

\author{Zaven Arzoumanian}
\affil{Astrophysics Science Division, NASA Goddard Space Flight Center, Greenbelt, MD 20771, USA}

\author{Peter Bult}
\affil{Astrophysics Science Division, NASA Goddard Space Flight Center, Greenbelt, MD 20771, USA}

\author{Deepto Chakrabarty}
\affil{MIT Kavli Institute for Astrophysics and Space Research, Massachusetts Institute of Technology, 70 Vassar Street, Cambridge, MA 02139, USA}

\author{Keith C.~Gendreau}
\affil{Astrophysics Science Division, NASA Goddard Space Flight Center, Greenbelt, MD 20771, USA}

\author[0000-0002-6089-6836]{Wynn C.~G.~Ho}
\affil{Department of Physics and Astronomy, Haverford College, 370 Lancaster Avenue, Haverford, PA 19041, USA}
\affil{Mathematical Sciences, Physics and Astronomy, and STAG Research Centre, University of Southampton, Southampton SO17 1BJ, UK}

\author[0000-0002-6789-2723]{Gaurava K.~Jaisawal} 
\affil{National Space Institute, Technical University of Denmark, 
  Elektrovej 327-328, DK-2800 Lyngby, Denmark}
  
\author{Christian Malacaria}
\affiliation{NASA Marshall Space Flight Center, NSSTC, 320 Sparkman Drive, Huntsville, AL 35805, USA} 
\affiliation{Universities Space Research Association, NSSTC, 320 Sparkman Drive, Huntsville, AL 35805, USA}

\author{M.~Coleman Miller}
\affiliation{Department of Astronomy and Joint Space-Science Institute, University of Maryland, College Park, MD 20742, USA}

\author[0000-0001-7681-5845]{Tod E.~Strohmayer}
\affiliation{Astrophysics Science Division and Joint Space-Science Institute, NASA Goddard Space Flight Center, Greenbelt, MD 20771, USA}

\author[0000-0002-4013-5650]{Michael T.~Wolff}
\affil{Space Science Division, Naval Research Laboratory, Washington, DC 20375, USA}

\author{Kent S.~Wood}
\affil{Praxis, resident at the Naval Research Laboratory, Washington, DC 20375, USA}

\author{Natalie A.~Webb}
\affil{IRAP, CNRS, 9 avenue du Colonel Roche, BP 44346, F-31028 Toulouse Cedex 4, France}
\affil{Universit\'{e} de Toulouse, CNES, UPS-OMP, F-31028 Toulouse, France.}

\author{Lucas Guillemot}
\affiliation{Laboratoire de Physique et Chimie de l'Environnement et de l'Espace, LPC2E, CNRS-Universit\'{e} d'Orl\'{e}ans, F-45071 Orl\'{e}ans, France}
\affiliation{Station de Radioastronomie de Nan\c{c}ay, Observatoire de Paris, CNRS/INSU, F-18330 Nan\c{c}ay, France}

\author{Ismael Cognard}
\affiliation{Laboratoire de Physique et Chimie de l'Environnement et de l'Espace, LPC2E, CNRS-Universit\'{e} d'Orl\'{e}ans, F-45071 Orl\'{e}ans, France}
\affiliation{Station de Radioastronomie de Nan\c{c}ay, Observatoire de Paris, CNRS/INSU, F-18330 Nan\c{c}ay, France}

\author{Gilles Theureau}
\affiliation{Laboratoire de Physique et Chimie de l'Environnement et de l'Espace, LPC2E, CNRS-Universit\'{e} d'Orl\'{e}ans, F-45071 Orl\'{e}ans, France}
\affiliation{Station de Radioastronomie de Nan\c{c}ay, Observatoire de Paris, CNRS/INSU, F-18330 Nan\c{c}ay, France}
\affiliation{LUTH, Observatoire de Paris, PSL Research University, CNRS, Universit\'{e} Paris Diderot, Sorbonne Paris Cit\'{e}, F-92195 Meudon, France}

\begin{abstract}

\nicer{} observed several rotation-powered millisecond pulsars to search for or confirm the presence of X-ray pulsations. When broad and sine-like, these pulsations may indicate thermal emission from hot polar caps at the magnetic poles on the neutron star surface.  We report confident detections ($\ge4.7\sigma$ after background filtering) of X-ray pulsations for five of the seven pulsars in our target sample: PSR~J0614$-$3329, PSR~J0636$+$5129, PSR~J0751$+$1807, PSR~J1012$+$5307, and PSR~J2241$-$5236, while PSR~J1552$+$5437 and PSR~J1744$-$1134 remain undetected. Of those, only PSR~J0751$+$1807 and PSR~J1012$+$5307 had pulsations previously detected at the 1.7$\sigma$ and almost 3$\sigma$ confidence levels, respectively, in \xmmlong{} data. All detected sources exhibit broad sine-like pulses, which are indicative of surface thermal radiation. As such, these MSPs are promising targets for future X-ray observations aimed at constraining the neutron star mass-radius relation and the dense matter equation of state using detailed pulse profile modeling. Furthermore, we find that three of the detected millisecond pulsars exhibit a significant phase offset between their X-ray and radio pulses.
\end{abstract}


\section{Introduction}
\label{sec:intro}

Millisecond pulsars (MSP) are old neutron stars (with characteristic ages $\tau\sim\ee{9}\yr$) named so based on their millisecond range rotation periods ($P\lesssim 25$ ms), the fastest known having $P\simeq1.4$ ms \citep{hessels06}.  In contrast, the bulk of the pulsar population is observed with spin periods in the $\sim$0.1--10\sec\ range. As they age, these pulsars evolve beyond the so-called ``pulsar death line'' \citep{sturrock71,hibschman01} where they stop emitting in the radio band.  Old pulsars spinning at millisecond periods must have experienced a ``recycling'' process of spin-up. Pulsar recycling occurs via the transfer of angular momentum due to the accretion of matter from a binary companion onto the neutron star (NS), during the low-mass X-ray binary phase \citep{bisnovatyi74,alpar82,radhakrishnan82}.  Once this recycling process ends, the magnetic field re-establishes the MSP magnetosphere, enabling the re-activation of the pulsed radio emission. Overall, MSPs are characterized by their large ages, exceptional rotational stability, and low magnetic fields ($B\sim\ee{8-9}\G$).

While strong evidence exists that low-mass X-ray binaries are the progenitors of MSPs \citep{wijnands98,papitto13}, not all MSPs are found in binary systems. This may be because the orbital modulations of a MSP have not yet been discovered \citep[possibly due to faintness, or long orbital periods, e.g.,][]{bassa16,kaplan16}, or because the NS is truly isolated.  Possible explanations invoke the ablation of the companion \citep[as in black-widow and redback systems,][]{chen13} or the disruption of the binary system (via stellar encounters in dense environments).  It has also been suggested that some isolated MSPs might have been formed by the direct collapse of a massive white dwarf \citep{freire14}.

X-ray pulsations in MSPs were first discovered in \rosat\ observations of PSR~J0437$-$4715 \citep{becker93}. These apparently thermal pulsations were ascribed to heating of the surface due to internal friction, or due to polar cap heating caused by magnetospheric return currents along open NS magnetic field lines \citep{harding01,harding02}.  Since this discovery, the X-ray emission from a handful of MSPs (mostly in globular clusters) has been identified as due to hot $\sim\ee{6}\K$ thermal (blackbody-like) emission from an area much smaller than the entire NS surface \citep[e.g.,][]{zavlin06,bogdanov06a,bogdanov11,forestell14,lee18}.  These small and hot polar caps at or close to the NS surface generate broad sine-like modulations with large pulsed fraction ($\sim$30--70\%) as the NS rotates. It was realized early on that the X-ray pulse profiles of MSPs could provide probes of the physical properties of NSs \citep[compactness, or mass-to-radius ratio,][]{pavlov97} when modeled with a realistic NS atmosphere \citep{zavlin96}.  This seminal work prompted detailed studies of the modeling of X-ray pulsations of MSPs \citep[e.g.,][]{bogdanov07,bogdanov08,bogdanov09}.

Measuring the compactness---and, in the best circumstances, the radius---of a NS is a crucial tool to determine the still-unknown equation of state (EOS) of dense nuclear matter. Obtaining constraints on the NS radius from the modeling of MSP pulse profiles provides a powerful method to discriminate between the numerous theoretical models describing dense nuclear matter \citep[for a recent review, see][]{lattimer16}.  Modeling MSP pulse profiles is a very promising technique that complements alternative methods making use of other classes of thermally emitting NSs \citep[e.g.,][]{heinke06a,heinke14,webb07,guillot13,guillot16b,ozel10,ozel16a,bogdanov16,steiner10,miller13,miller16,steiner13,steiner18,baillot19}, and methods exploiting the gravitational wave signals from NS-NS mergers as demonstrated with the GW~170817 event \citep[e.g.,][]{abbott17,abbott18,de18}.

The \emph{Neutron star Interior Composition Explorer} (\nicer; \citealt{gendreau17}) was designed to fully exploit the pulse profile modeling technique which requires data with high signal-to-noise ratios (S/N), the main scientific goal of the \nicer{} mission being to constrain the radii of NSs from observations of a handful of MSPs. An additional science goal of the mission is to study the X-ray flux modulations of pulsars and discover new sources of X-ray pulsations \citep{2018IAUS..337..187R}. This includes searching for X-ray pulsations from NSs that would be suitable for pulse profile modeling with \nicer{} or with future missions (e.g., \textit{STROBE-X}, \citealt{ray19a}; \textit{eXTP}, \citealt{watts19}; or the \textit{Athena X-ray Observatory}, \citealt{nandra13}).  Prior to the launch of \nicer{}, only four thermally-dominated MSPs were known to exhibit highly significant X-ray pulsations: PSR~J0437$-$4715 \citep{becker99,zavlin02b,bogdanov13}, PSR~J0030$+$0451 \citep{becker02,bogdanov08}, PSR~J2124$-$3358 \citep{becker99,zavlin06,bogdanov08} and PSR~J1024$-$0719 \citep{zavlin06}. A few other MSPs showed tentative or marginal detections: PSR~J0751$+$1807 at $1.7\sigma$ and PSR~J1012$+$5307 at almost $3\sigma$ \citep{webb04b}; and PSR~J1614$-$2230, at the  $4\sigma$ level \citep{pancrazi12}. 

Most radio MSPs hosted in globular clusters are found to be positionally coincident with soft, thermal X-ray sources \citep[e.g.,][]{bogdanov06a,bogdanov11,forestell14}, whose spectra suggests that their emission also arises from the heated magnetic polar caps of the NS. A long \chandra\ HRC observation of 47~Tuc yielded $4\sigma$ pulsation detections for three MSPs hosted by this cluster \citep{cameron07}.  Some MSPs with non-thermal emission also exhibit pulsations in the X-ray band, although they tend to be characterized by short--duty-cycle pulse profiles with hard non-thermal spectra (e.g., PSR~B1821$-$24 in the globular cluster M28, PSR~B1937$+$21, \citealt{gotthelf17,deneva19}), likely caused by magnetospheric emission. Others, such as PSR~J0218$+$4232, have a hard non-thermal spectrum with moderately broad X-ray profiles \citealt{webb04a,deneva19}.

Aside from the thermally-emitting MSPs mentioned above, those in globular clusters, and the non-thermal emitters, a number of relatively nearby MSPs show apparently thermal X-ray emission, but were not previously observed in a mode permitting searches for pulsations and/or with sufficient exposures. These include PSR~J0023+0923 (a black-widow with a very soft X-ray spectrum; \citealt{gentile14}), PSR~J0034$-$0534 \citep{zavlin06}, PSR~B1257$+$12 (the pulsar with planets; \citealt{pavlov07}), PSR~J1400$-$1431 (a very nearby but very X-ray faint MSP; \citealt{swiggum17}), and PSR~J1909$-$3744 \citep{kargaltsev12,webb19}.  A complete list of X-ray emitting MSPs in the Galactic field has been compiled by \cite{lee18}.

No systematic study exists on the offsets between the radio and X-ray pulses of rotation-powered MSPs, in part because of the challenging aspects of properly aligning the pulse profiles.  For those MSPs with thermal emission, originating from the footprints of the magnetic field at the NS surface, one naturally expects the X-ray and radio profiles to be aligned. This is indeed observed for a handful of MSPs: PSR~J0437$-$4715 \citep{bogdanov13}, PSR~J0030$+$0451 \citep{bilous19}, or PSR~J1231$-$1411 \citep{ray19b}. However, some exceptions include PSR~J1614$-$2230 \citep{pancrazi12} or PSR~J2124$-$3358 \citep{becker99}, although for the later, the $1\sigma$ uncertainty is 1\,ms, i.e., $\sim$20\% of the total spin period.  For non-thermal MSPs, for which the emission originates in the magnetosphere, some have been observed with near perfect alignment (PSR~B1821$-$24 or PSR~B1937$+$21, \citealt{deneva19}), possibly suggesting a similar location for the origin of the radio and X-ray emission.  Others, however, show a significant offset, such as PSR~J0218$+$4232, in which the broad double-peaked radio pulse lies halfway between the two moderately broad X-ray pulses \citep{deneva19}.

A selection of nearby MSPs in the field of the Galaxy with known thermal emission, but no prior firm detection of X-ray pulsations, have been targeted by \nicer{} to detect or confirm the presence of polar-cap emission causing pulsations.  Here we report on the results from these observations and the discovery of pulsations in some of them. Section~\ref{sec:msp} briefly presents the targets and some of their basic properties. Section~\ref{sec:obs} describes the observations performed, the data reduction and search method. The results are presented in Section~\ref{sec:results} and are followed by a brief discussion on the properties of the detected pulsations and conclusions in Section~\ref{sec:discussion}.

\section{Targets}
\label{sec:msp}

For this investigation, we considered MSPs that had no previous securely identified X-ray pulsations. This included targets with only marginal detections of pulsations.  The basic properties of these MSPs are summarized in Table~\ref{tab:targets}, and we provide below an overview of some of their characteristics and details about prior X-ray observations. Other MSPs with known pulsations and observed by \nicer\ for EoS determination purposes are presented in \cite{bogdanov19a}.

\subsection{PSR~J0614$-$3329} 
The 3.1 ms spin period of PSR~J0614$-$3329 was discovered by the Robert C.~Byrd Green Bank Telescope (GBT) from a source detected in $\gamma$-rays with the \textit{Fermi} Large Area Telescope \citep[LAT;][]{ransom11} but with no known counterpart at any other wavelength. With a period derivative of $\dot{P}\approx1.75\tee{-20}\unit{\sec\persec}$, this pulsar's surface dipolar magnetic field is $B\approx2.38\tee{8}\G$ \citep{ransom11}.  The orbital period of 53.6\,d was also measured from radio timing.  The companion star, tentatively a helium white dwarf, was identified from its optical colors in $g$, $r$, and $i$ band Gemini observations, despite the pulsar's proximity to a background galaxy \citep[$\sim 10\arcsec$, ][]{testa15}.  Finally, X-ray data obtained with the \textit{Neil Gehrels Swift Observatory} indicated a likely thermal X-ray spectrum with a blackbody temperature $kT_{\rm BB}=0.23\pm0.05\keV$ \citep{ransom11}.

\begin{deluxetable}{lrDDrcr}
\tablecaption{Selection of nearby MSPs targeted by \nicer \label{tab:targets}}
\tablehead{
\colhead{Pulsar} & \colhead{$P$} & \multicolumn2c{$\dot{P}$} & \multicolumn2c{$P_{\rm orb}$} & \colhead{$D$} &\colhead{$M$} & \colhead{Timing} \\
  & \colhead{(ms)} & \multicolumn2c{($\times 10^{-21}$)} &\multicolumn2c{(d)}  & \colhead{(pc)} & \colhead{($\Msun$)} & \colhead{Solution}  }
\decimals
\startdata
PSR~J0614$-$3329 & 3.10 & 17.5  & 53.6    & $\sim$2690          & unknown              & NRT \\
PSR~J0636$+$5129   & 2.87 & 3.38  & 0.066   & $203\ud{27}{21}$    & unknown              & NRT   \\
PSR~J0740$+$6620   & 2.88 & 12.2  & 4.77    & $400\ud{200}{100}$  & 2.17\ud{0.11}{0.10}  & (1) \\
PSR~J0751$+$1807   & 3.48 & 7.79  & 0.25    & $1070\ud{240}{170}$ & 1.64\ppm0.15         & NRT \\
PSR~J1012$+$5307   & 5.26 & 17.1  & 0.6     & 907\ppm131          & 1.83\ppm0.11         & NRT \\
PSR~J1231$-$1411 & 3.68 & 22.8  & 1.86    & $\sim$420           & unknown              & (2) \\
PSR~J1552$+$5437   & 2.43 & 2.80  & ...     & $\sim$2600          & \nodata              & \fermi/LOFAR \\
PSR~J1614$-$2230 & 3.15 & 9.624 & 8.7     & $670\ud{50}{40}$    & 1.908\ppm0.016       & (1) \\
PSR~J1744$-$1134 & 4.08 & 8.935 & ...     & $440\ud{20}{20}$    & \nodata              & NRT\\
PSR~J2241$-$5236 & 2.19 & 6.64  & 0.15    & $\sim$960           & unknown              & Parkes \\
\enddata
\tablenotetext{1}{\nicer\ observations presented in \cite{arzoumanian19}}
\tablenotetext{2}{\nicer\ observations presented in \cite{ray19b}.}
\tablecomments{Distances with quoted uncertainties are based on measurements of parallax from radio timing. For the rest, a dispersion measure (DM)-based distance estimate is given \citep{yao17}. The $\dot{P}$ value reported are not corrected for the Shklovskii effect. ``NRT'' stands for the \nrt.}
\end{deluxetable}

\subsection{PSR~J0636$+$5129}
The Green Bank Northern Celestial Cap (GBNCC) survey for pulsars discovered this 2.87\msec\ pulsar in a 95.6\,minute binary orbit \citep{stovall14}.  Based on its orbital properties, the system was initially classified as a black widow binary, although some features typically seen in black widow pulsars are lacking (e.g., eclipses and dispersion measure variations; \citealt{stovall14}). The binary companion was recently discovered as a magnitude $r\approx24$ low-mass ($M<0.02\msun$) companion star. As no radial velocity information was available, no constraints on the pulsar mass were obtained \citep{draghis18}. \textit{XMM-Newton} data of PSR~J0636$+$5129 showed a thermal spectrum, with $kT_{\rm BB}=0.18\pm0.03\keV$. Although a purely non-thermal model could not be formally excluded, the resulting power-law photon index was unusually soft, $\Gamma=5\ud{5}{1}$, indicating a likely thermal spectrum \citep{spiewak16}.

\subsection{PSR~J0751$+$1807}
PSR~J0751$+$1807, a MSP with $P=3.48\msec$, was discovered by the Arecibo Observatory in a search targeting unidentified EGRET sources \citep{lundgren95}. The MSP is in a 6\,hr binary system with a helium white dwarf companion. Initial mass measurements via radio timing detection of orbital decay and Shapiro delay resulted in a NS mass of $2.1\pm0.2\msun$ \citep{nice05}.  However, more recent radio timing observations found a significantly different NS mass of $1.64\pm0.15\msun$ and a companion mass of $0.16\pm0.01\msun$ \citep{desvignes16}.  The reported parallax distance of this MSP is $D=1.07\ud{0.24}{0.17}\kpc$.  Finally, its spin down $\dot{P}\approx0.8\tee{-20}\unit{\sec\persec}$ implies a surface dipole magnetic field strength of $B\approx1.7\tee{8}\G$.  In the X-ray waveband, the \xmm\ spectrum is well described by a power law with photon index $\Gamma=1.6\pm0.2$ \citep{webb04b}, but not by a single blackbody, which suggests the possibility of a non-thermally emitting MSP.  These data also showed a hint of pulsations at the $1.7\sigma$ confidence level. 

\subsection{PSR~J1012$+$5307}
PSR~J1012$+$5307 was discovered with the Lovell radio telescope at Jodrell Bank Observatory \citep{nicastro95}.  This 5.26\msec\ pulsar is in an orbit with a low-mass helium white dwarf \citep{vankerkwijk96}.  While some orbital parameters (e.g., $P_{\rm orb}\approx0.6\unit{d}$) were measured in the discovery observation, they have since then been refined \citep{fonseca16}. Using the radial velocity curve of the companion, as well as the companion mass (determined via optical spectroscopy), \cite{callanan98} estimated the pulsar mass to be $1.64\pm0.22\msun$, and a more recent estimate resulted in $1.83\pm0.11\msun$ \citep{antoniadis16}. Having an independent mass measurement makes this pulsar a potentially interesting \nicer{} target for radius measurements.  The X-ray spectrum of PSR~J1012$+$5307, obtained with \textit{XMM-Newton}, is consistent with both absorbed power law ($\Gamma\sim1.8$) and absorbed blackbody models \citep{webb04b}.  If the emission is thermal, the blackbody temperature, $kT_{\rm BB}=0.26\pm0.04\keV$, is consistent with that of other similar MSPs. \cite{webb04b} also reported a $3\sigma$ detection of X-ray pulsations at the spin frequency of the pulsar. The distance to this pulsar has recently been updated by identifying the optical counterpart in the \gaia-DR2 catalog, resulting in a distance measurement of $907\pm131\pc$ when combining the \gaia\ parallax with various other measurements \citep{mingarelli18}.

\subsection{PSR~J1552$+$5437}
PSR~J1552$+$5437 was recently discovered in the radio band with the Low-Frequency Array (LOFAR, \citealt{vanhaarlem13}) in an unassociated \textit{Fermi}-LAT $\gamma$-ray source.  This $P=2.43\msec$ pulsar was the first MSP to be discovered at low radio frequencies \citep[115--150\,MHz,][]{pleunis17}, and its timing solution was used to confirm the presence of pulsations at the same spin period in the original \fermi\ LAT source 3FGL~J1553.1+5437.  No X-ray counterpart was detected in a short 2.9\ksec\ observation with the \textit{Neil Gehrels Swift Observatory}. No optical counterpart is known for this pulsar.

\subsection{PSR~J1744$-$1134} 
Discovered in the Parkes 436\,MHz survey of the southern sky \citep{manchester96}, this nearby $P=4.08\msec$ pulsar is now routinely observed for pulsar timing array purposes \citep[e.g.,][]{reardon16,arzoumanian18}.  The X-ray counterpart of PSR~J1744$-$1134 was discovered in \rosat\ data \citep{becker99}. It has since then been observed by \chandra, with the X-ray spectrum likely indicating thermal emission with black-body temperature of $kT_{\rm BB}\sim0.3\keV$ \citep{marelli11}.   No optical counterpart has been found despite deep searches, setting an upper limit of $V<26.3$ \citep{sutaria03}.

\subsection{PSR~J2241$-$5236} 
PSR~J2241$-$5236 is another pulsar discovered among \fermi\ LAT unidentified $\gamma$-ray sources \citep{keith11}. It has a spin period of 2.19\msec. It is classified as a black-widow pulsar with intra-binary shock, in a 3.5\,hr orbit with a $M>0.012\msun$ companion, at a dispersion measure distance of 0.960\kpc\ \citep{yao17}. A \chandra\ observation of PSR~J2241$-$5236 identified a soft X-ray source with a spectrum described by two black bodies with temperatures of $kT_{\rm BB}\sim 0.07\keV$ and $\sim 0.26\keV$ \citep{keith11}.

\subsection{Other thermally-emitting MSPs}
The following other MSPs were also observed with \nicer, but their results are not reported in the present paper:
\begin{itemize}
    \item PSR~J1231$-$1411, a 3.68\msec\ pulsar, was discovered in a search campaign of unassociated \textit{Fermi}-LAT sources with the GBT \citep{ransom11}. With a 0.2--12\keV\ flux of $1.9\tee{-13}\cgsflux$, it is the third brightest thermally-emitting MSP and thus a target well suited for NS compactness measurement with \nicer.  The pulsar is in a 1.86-d orbit with a low-mass white dwarf \citep{ransom11}, possibly associated with a magnitude $g=25.4$ optical counterpart \citep{testa15}.  X-ray pulsations from PSR~J1231$-$1411 were detected, but are reported in detail elsewhere \citep{ray19b,bogdanov19a}.
    \item PSR~J1614$-$2230 was discovered in a Parkes radio search targeting unidentified EGRET sources \citep{hessels05,crawford06}. It is a MSP with $P=3.15\msec$ bound to a massive white dwarf companion in a $P_{b}=8.7\unit{d}$ orbit. The pulsar is particularly important for the \nicer\ mission because it is one of the most massive NSs known \citep{miller16}, with $M=1.908\pm0.016\msun$ \citep{arzoumanian18}. X-ray pulsations from this pulsar were detected with \nicer, but are reported in full detail elsewhere \citep{arzoumanian19}, confirming a $\sim4\sigma$ detection in \xmm\ data \citep{pancrazi12}.
    \item PSR~J0740+6620 is another GBNCC-discovered pulsar, in a 4.77\unit{d} orbit with a white-dwarf companion \citep{stovall14,lynch18,beronya19}. An intense timing campaign with the GBT permitted detection of a Shapiro-delay signal, resulting in a pulsar mass $2.14\ud{0.10}{0.09}\msun$ \citep{cromartie19}. Although with somewhat large uncertainties, the high NS mass puts this pulsar on a par with PSR~J1614$-$2230 for dense matter EOS constraints \citep{miller19}. The \nicer\ observations of this pulsar are also reported in \cite{arzoumanian19}.
\end{itemize}

\clearpage
\section{Observations and data reduction}
\label{sec:obs}

\subsection{\nicer{} X-ray Timing Instrument, data filtering and pulse phase assignments}

We observed the seven MSPs listed in Section~\ref{sec:msp} using \nicer. Table~\ref{tab:obs} presents these data, giving the ObsIDs used, the total exposure accumulated, and the total exposure available after filtering. We processed all the data available until 2019 March 23 with {\tt HEASOFT} v6.23 and the \nicer{} specific package {\tt NICERDAS} v3.  We filtered the raw data with the standard criteria:
\begin{itemize}
    \setlength\itemsep{0.2cm} 
    \item pointing offset is $<0.015\deg$ from the source,
    \item pointing $>20\deg$ from the Earth limb ($>30\deg$ in the case of a bright Earth),
    \item excluding South Atlantic Anomaly passages,
    \item selecting events in the 0.25--12 keV energy range. 
\end{itemize}

\begin{deluxetable}{lrrcc}
\tablecaption{\textit{NICER} observations of selected millisecond pulsars\label{tab:obs}}
\tablehead{
\colhead{Targeted} & \colhead{ObsID} & \colhead{Raw }  & \colhead{Std. Filt.} & \colhead{Opt. Filt.} \\
\colhead{Pulsar} & \colhead{range} & \colhead{ Exp. (ks)}  & \colhead{Total Exp. (ks)} & \colhead{Total Exp. (ks)}
}
\startdata
PSR~J0614$-$3329 &   0030050101 -- 0030050110 &  82 & \multirow{3}{*}{188} & \multirow{3}{*}{181} \\
				& 	1030050101 -- 1030050158 & 151 & \\
				& 	2030050101 -- 2030050103 &  29 & \\
\hline
PSR~J0636$+$5129  &   1030070101 -- 1030070197 & 403 & \multirow{2}{*}{399} & \multirow{2}{*}{347} \\
				& 	2030070101 -- 2030070107 &  40 & \\
\hline
PSR~J0751$+$1807  &   1060030101 -- 1060030217 & 462 & \multirow{2}{*}{443} &\multirow{2}{*}{424} \\
				& 	2060030201 -- 2060030211 &  51 & \\
\hline
PSR~J1012$+$5307  &   0070040101 -- 0070040113 &  45 & \multirow{3}{*}{590} &\multirow{3}{*}{548}\\
 				&   1070040101 -- 1070040215 & 570 & \\
 				&   2070040201 -- 2070040212 &  54 & \\
\hline
PSR~J1552$+$5437  &   1033180102 -- 1033180116 & 75  & 72   & \nodata\\
\hline
PSR~J1744$-$1134 &   0030160101 -- 0030160109 & 16  & \multirow{3}{*}{71} & \multirow{3}{*}{\nodata}\\
				&   1030160101 -- 1030160125 & 62  & \\
				&   2030160101 -- 2030160106 &  5  & \\
\hline
PSR~J2241$-$5236 &   1031010101 -- 1031010182 & 254 & \multirow{2}{*}{113} & \multirow{2}{*}{100}\\
				&   2031010101               &   3 & \\
\enddata
\tablecomments{These ObsIDs include all observations acquired from the beginning of the mission (2017 July) through 2019 March 23. The exposure time columns report, from left to right, the total duration of data collection (``Raw''), the exposure time after the standard filtering described in Section~\ref{sec:obs} (``Std. Filt.''), and the final exposure time resulting from the GTI-sorting optimization (``Opt. Filt.''). For J1552$+$5437 and J1744$-$1134, no optimized exposure time is given since no detection is claimed (see Section~\ref{sec:results}).} 
\end{deluxetable}

Additional filtering was necessary to ensure minimal contamination from non-astrophysical background. First, we excluded detectors with \texttt{DET\_ID} 14, 34, and 54---these are known to be more sensitive to optical loading than others, complicating analyses for soft sources such as our MSPs. Then, we used a filtering criterion based on the rate of ``overshoots'' (representing large deposition of energy within a detector, and defined by the housekeeping parameter \texttt{FPM\_OVERONLY\_COUNT}) and on the magnetic cutoff rigidity (\texttt{COR\_SAX}, in GeV/$c$). An empirical relation was found to minimize periods of high background, even at low cutoff rigidities, in order to maximize the exposure. For each pulsar, we excluded observed time intervals with the conditions\footnote{The filtering criteria on \texttt{FPM\_OVERONLY\_COUNT} are adopted into the standard {\tt NICERDAS} pipeline as of version 6.0 (\url{https://heasarc.gsfc.nasa.gov/docs/nicer/nicer_analysis.html}).}: 
\begin{align*} 
 \texttt{FPM\_OVERONLY\_COUNT} & > (1.52\times\texttt{COR\_SAX}^{-0.633}) \\
 \text{and }  \texttt{FPM\_OVERONLY\_COUNT} & > 1.0
\end{align*}
Some pulsars required slightly different filtering options. For example, a portion of the exposure for PSR~J1552$+$5437 was acquired with a pointing offset $>0.015\deg$, and therefore this condition was relaxed to include pointing offset $<0.1\deg$. When necessary, these additional criteria are described for each pulsar in Section~\ref{sec:results}.

Finally, the pulse phase of each photon was computed using the pulsar timing analysis software PINT (task {\tt photonphase}), given the ephemeris for the pulsar of interest (see Appendix~\ref{app:ephem}). Note that {\tt photonphase} computes the transformation from the Terrestrial Time (TT) standard used for time tagging of \nicer{} events to Barycentric Dynamical Time (TDB) using the \nicer{} orbit files (provided as one of the standard products for each ObsID) and the pulsar astrometric parameters (position, proper motion, and parallax) provided in the epheremides---only the Solar System Ephemeris has to be specified (DE421 or DE436).  The H-test is then employed to quantify the presence of pulsations \citep{dejager89,dejager10}, either in the full band, or in the soft X-ray band (0.25--2.0\keV), setting a $3\sigma$ single-trial limit to claim the detection of pulsations. We report the detections and non-detections of pulsations in Section~\ref{sec:results}.  Despite our stringent filtering described above, the background, variable throughout the ObsIDs, may still contaminate significantly the pulse profiles of the detected pulsars.  We describe in the next section a method of GTI selection to minimize background contamination and provide cleaner pulse profiles.

\subsection{Good time interval (GTI) sorting method}
\label{sec:gtisort}

\begin{figure}[t]
\begin{center}
\includegraphics[width=0.8\textwidth]{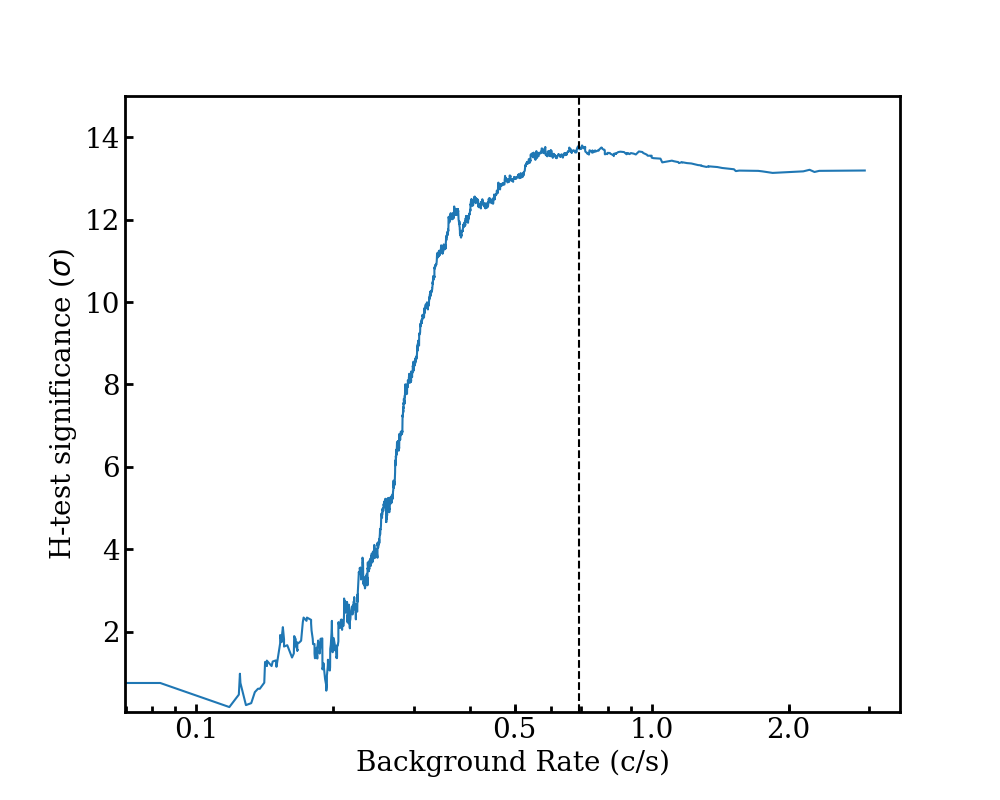}
\caption{Evolution of the pulsation detection significance (units of $\sigma$) as a function of the GTIs sorted by total count rate, equivalent to the cumulative exposure time (totaling 181\ksec), for PSR~J0614$-$3329. The dashed vertical line indicates the optimal GTI cut that maximizes the detection significance.  In other words, adding GTIs to the left of this line increases the detection of pulsations, while adding high-background GTIs, to the right of the line, decreases the detection significance. 
}
\label{fig:GTIsort}
\end{center}
\end{figure}

Because each \nicer{} exposure occurs under different observing conditions (pointing with respect to the Sun, Earth, or Moon, location of the ISS, space weather, etc.), the amount of non-astrophysical background can vary greatly between ObsIDs, and even within single ObsIDs. Faint sources, such as the MSPs studied in this paper, are most affected by background: GTIs with low background improve the S/N, while those with high background dilute the pulsations and decrease the S/N.  To minimize the background contamination and maximize the S/N, we therefore only consider the GTIs with low background. This is done by splitting the GTIs into small segments of at least 10\,s (to have a good estimate on the background count rate) and at most 100\,s. These GTIs are then ordered by their corresponding total count rate, which we use as a proxy for the background rate since the MSP contribution to the total rate is of the order of a few percent and is constant.  We employ the H-test to quantify the significance of the pulsations. We do so cumulatively on the sorted GTIs (from the lowest to the highest count rate).  This is illustrated in Figure~\ref{fig:GTIsort} for PSR~J0614$-$3329, which shows the cumulative H-test significance as a function of GTIs with increasing background count rates. The H-test significance first increases as low-background GTIs are added and then stabilizes or decreases when higher background GTIs are being included. This methods permits finding the optimal cut (indicated by the vertical line in Figure~\ref{fig:GTIsort}) that maximizes the H-test significance by excluding GTIs with the highest background rates.

To further increase the pulse profiles S/N, i.e., reducing the background contributions, additional improvement can be achieved by optimizing the selected energy range. In some GTIs, optical loading may cause a sharp increase of the count rate below $\sim 0.4\keV$. Excluding part or all of the 0.25--0.4\keV\ energy range may therefore further increase the detection significance of the pulsations.  Similarly, since the thermal spectra of MSPs are expected to drop rapidly above $\sim$1.5 or 2\keV, an optimization of the high-energy cut-off can minimize the background contribution to the pulse profiles. Here, we perform a grid search for the low- and high-energy cut-offs (with ranges 0.25--1.0\keV\ and 0.9--3.0\keV, respectively) and the resulting optimal energy ranges are reported in Table~\ref{tab:res}.

\subsection{\nrt\ pulsar timing data}

Timing solutions for all pulsars, except PSR~J1552$+$5437 and PSR~J2241$-$5236, have been constructed by analyzing pulsar observations made with the Nan\c{c}ay Ultimate Pulsar Processing Instrument (NUPPI) in operation at the \nrt\ (NRT) since August 2011. In these observations, 512\,MHz of frequency bandwidth are recorded in the form of 128 channels of 4\,MHz each, which are coherently de-dispersed in real time and phase-folded at the expected topocentric periods of the observed pulsars \citep[see][for additional details on pulsar timing observations with the NUPPI backend]{guillemot16}. We considered timing observations made at 1.4\,GHz, which represent the bulk of pulsar observations with NRT. For each pulsar, we used all of the available NUPPI data through 2019 June. 

Data reduction steps were performed using the \textsc{Psrchive} software library \citep{hotan04}. We cleaned the data of radio frequency interference and calibrated the observations in polarization using the \textsc{SingleAxis} method. High-S/N profiles were built by summing up the 10 best detections of each pulsar, and times of arrival (TOAs) were extracted by cross-correlating the observations with smoothed versions of the summed profiles. For each NUPPI observation, we formed one TOA per 128 MHz of bandwidth. The TOAs at multiple frequencies enabled us to track potential variations of the dispersion measure (DM). We used the TOA datasets and the \textsc{Tempo2} pulsar timing package \citep{hobbs06} to build timing solutions for each pulsar, fitting for their astrometric, rotational, and DM parameters, as well as orbital parameters for those pulsars in binary systems. The timing solutions, presented in Appendix~\ref{app:ephem}, obtained with this procedure describe the TOAs appropriately. We note that these timing solutions are sufficient for the purpose of calculating the phases of X-ray photons, as done in the present paper, over the span of NICER's data collection, but they might not be the most optimal solution for long-term timing of these pulsars.

\subsection{Parkes pulsar timing data}

Timing solutions for PSR~J2241$-$5236 made use of data collected with the Parkes radio telescope, primarily for the Parkes Pulsar Timing Array (project P456).  The observing strategy is described in \citet{reardon16}, and the data reduction by \citet{kerr18}, but we summarize them here.  Observations of 64\,min duration are performed with approximately a three-week cadence in three radio frequency bands, typically centered on 732\,MHz, 1369\,MHz, and 3100\,MHz.  Down-converted voltages are digitized, converted to spectra via a polyphase filterbank, and autocorrelated to form Stokes parameters.  These multi-channel time series are folded at the known pulsar spin periods into pulse profiles and integrated for 64\,s before output for archiving.  Observations of a pulsed noise diode enable gain and polarization calibration.  Data are reduced using  \textsc{Psrchive} and TOAs extracted using an analytic profile.  As described above, timing solutions are prepared with \textsc{Tempo2}.  To enable absolute alignment of the radio and X-ray profiles, we selected a single high-S/N TOA to serve as the reference epoch (the \texttt{TZRMJD} parameter; see Section~\ref{sec:phase}).  The timing solution makes use of radio data collected between 2010 Feb 9 and 2018 Apr 22, requiring extrapolation to fold \nicer{} data acquired outside of this epoch.  We verified the timing solution on ad hoc data acquired through 2018 Nov 10 and observed no substantial deviations of these pulse arrival times from the predicted phase.

\section{Results}
\label{sec:results}

The results of the pulsation searches are presented in Table~\ref{tab:res}, which contains the single-trial detection significance, the H-test significance after GTI optimization, and the optimal energy range for each observed MSP.  The optimization of the energy range permitted finding the low-energy cut, which depends on how much optical-loading noise is present, as well as the high-energy cut, which depends primarily on the brightness of the MSP (and its spectral shape, i.e., surface temperature). Five out of the seven pulsars show pulsations in the soft X-ray band, with a single-trial significance (before GTI optimization) $>4.3\sigma$, except for PSR~J1012$+$5307 (at $3\sigma$).  With the GTI optimization described above, the H-test significance for the detected pulsars are all $>4.7\sigma$.  The two non-detected pulsar have single-trial H-test significance of 1.5$\sigma$ and 2.5$\sigma$ for PSR~J1552$+$5437 and PSR~J1744$-$1134, respectively. 

\begin{deluxetable}{lDDcc}
\tabletypesize{\normalsize}
\tablecaption{Results of the pulsation searches \label{tab:res}}
\tablehead{
\colhead{Pulsar} & \multicolumn2c{single-trial H-test} & \multicolumn2c{GTI opt. H-test} & \colhead{Energy range}& \colhead{Pulsed} \\ 
  & \multicolumn2c{significance\tablenotemark{a}} & \multicolumn2c{significance\tablenotemark{b}} & \colhead{(keV)} & \colhead{count rate\tablenotemark{c} (s$^{-1}$)} 
}
\decimals
\startdata
PSR~J0614$-$3329 & $11.0\sigma$ & $14.0\sigma$ & 0.33--1.43 & 0.027\ppm0.002 \\ 
PSR~J0636$+$5129 & $4.3\sigma$  & $5.5\sigma$  & 0.27--0.91 & 0.008\ppm0.001 \\ 
PSR~J0751$+$1807 & $6.5\sigma$  & $8.5\sigma$   & 0.32--1.82 & 0.014\ppm0.001 \\ 
PSR~J1012$+$5307 & $3.0\sigma$  & $4.7\sigma$   & 0.31--1.94 & 0.007\ppm0.001 \\ 
PSR~J1552$+$5437 & $1.5\sigma$  & \multicolumn{4}{c}{\emph{No detection}} \\
PSR~J1744$-$1134 & $2.5\sigma$  & \multicolumn{4}{c}{\emph{No detection}} \\
PSR~J2241$-$5236 & $6.3\sigma$  & $7.4\sigma$   & 0.41--1.14 & 0.020\ppm0.002 \\ 
\enddata
\tablenotetext{{\rm a}}{Single-trial H-test significance in the 0.25--2.0\keV\ energy range.}
\tablenotetext{{\rm b}}{H-test significance after optimization of the GTI and energy band (see Section~\ref{sec:discussion} for a short discussion of the number of trials).}
\tablenotetext{{\rm c}}{ Pulsed count rates are provided in the optimal energy range provided for each pulsar. }
\tablecomments{The results for PSR~J1231$-$1411, PSR~J0740+6620, and PSR~J1614$-$2230 are reported elsewhere (\citealt{ray19b} and \citealt{arzoumanian19}).}
\end{deluxetable}

\subsection{PSR~J0614$-$3329}

\begin{figure*}[t]
\begin{center}
\includegraphics[width=0.5\textwidth]{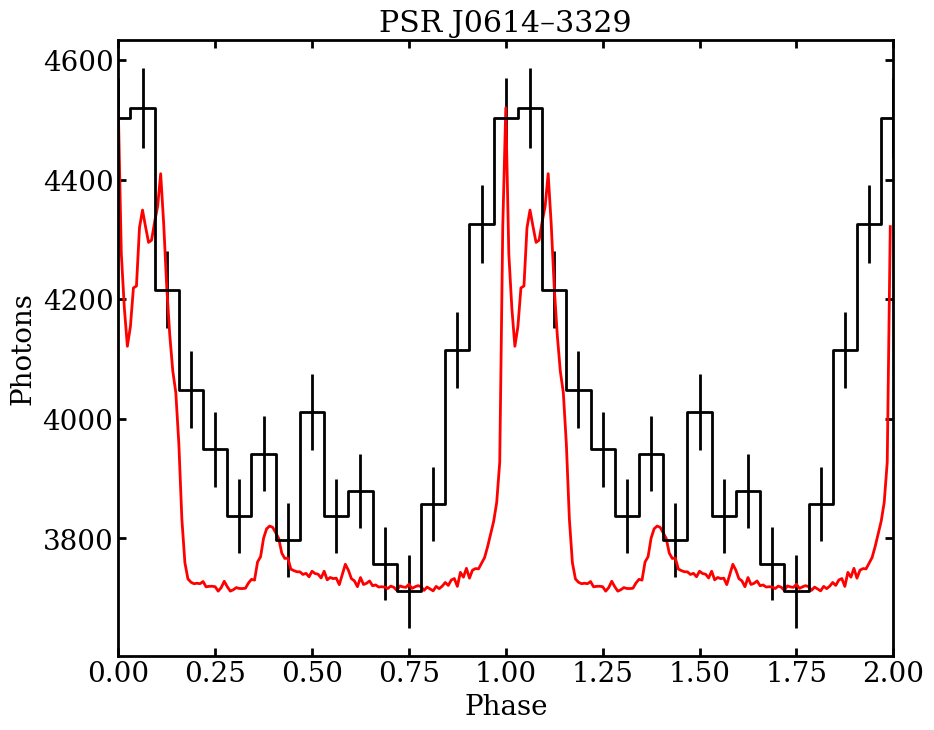}~
\includegraphics[width=0.5\textwidth]{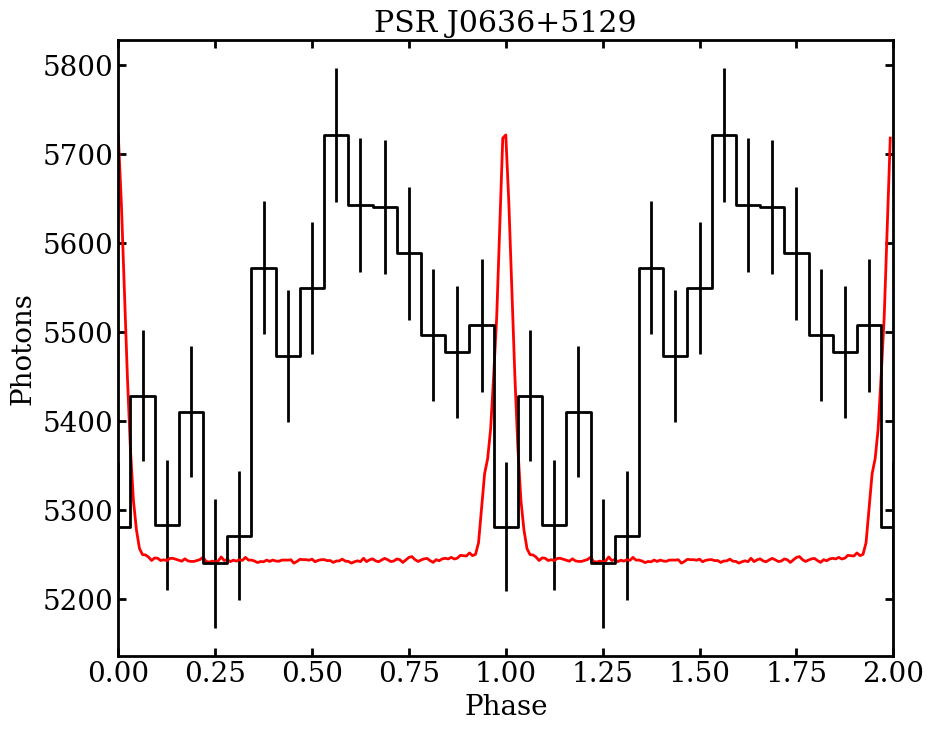}
\includegraphics[width=0.5\textwidth]{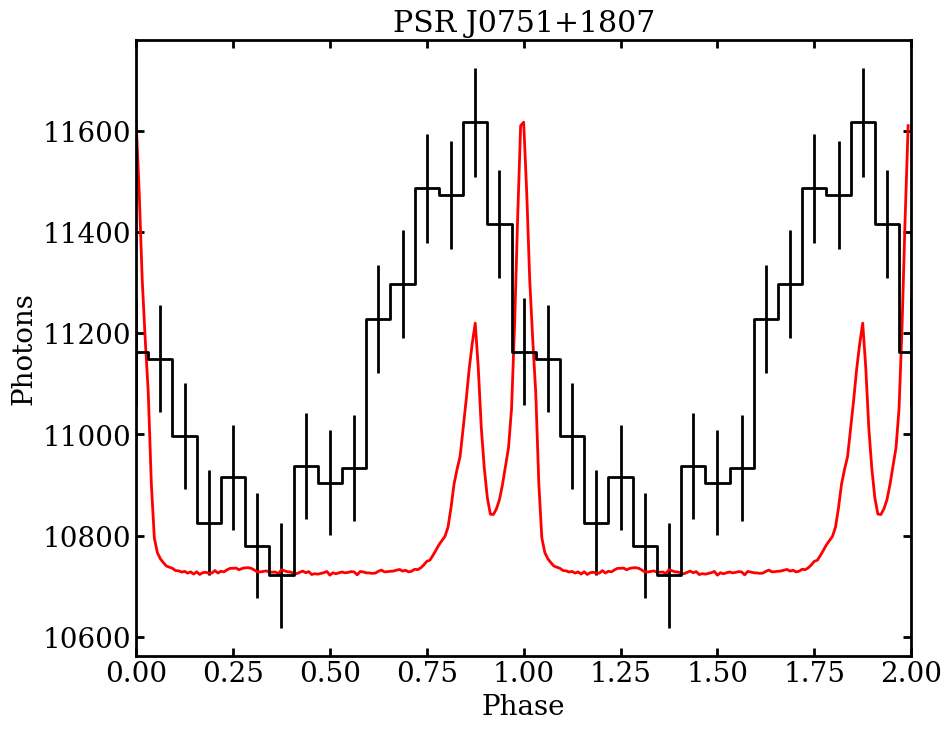}~
\includegraphics[width=0.5\textwidth]{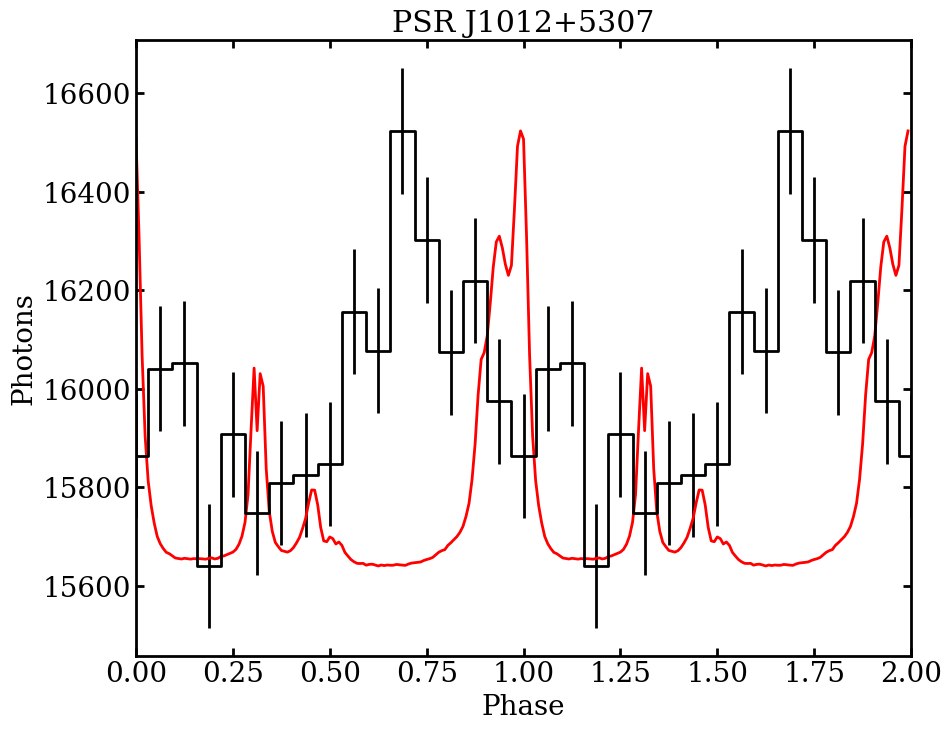}
\includegraphics[width=0.5\textwidth]{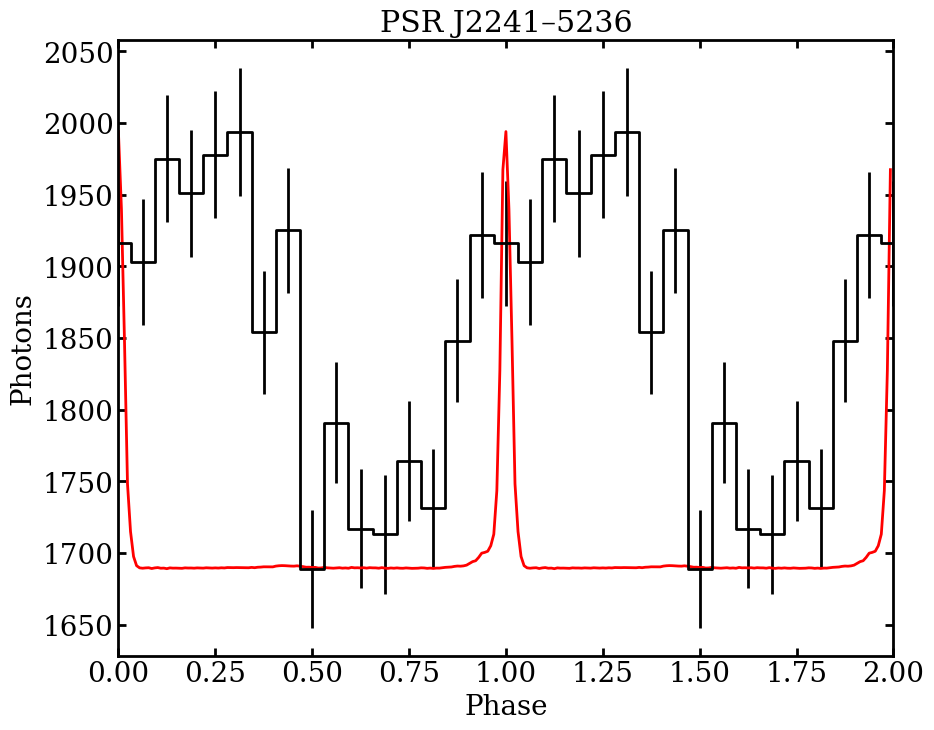}
\caption{\textit{NICER} pulse profiles (black, 16 bins per phase) for the five pulsars with detected pulsations (obtained in their respective optimal energy ranges, see Table~\ref{tab:res}), together with their radio pulse profile at 1.4\,GHz (red) from the \textit{Nan\c{c}ay} or \textit{Parkes} radio telescopes. The ephemerides used for folding are presented in Tables~\ref{tab:ephem0614} through  \ref{tab:ephem2241}, respectively. Two rotational cycles are shown for clarity.
}
\label{fig:profiles}
\end{center}
\end{figure*}

For this pulsar, a simple phase-folding of the events with a recent ephemeris (Table~\ref{tab:ephem0614}) in the full \nicer\ energy range resulted in a significant $7.5\sigma$ single-trial detection of the pulsations, and $11\sigma$ when selecting events in the soft (0.25--2.0\keV) band. Optimization of the GTIs (see Section~\ref{sec:gtisort}) improved the significance to $11.4\sigma$. A grid search in energy concluded that the 0.33--1.43\keV\ energy range is optimal, resulting in a $14\sigma$ significance, for a total exposure time of 181\ksec. The pulse profile is shown in Figure~\ref{fig:profiles}, together with the NRT pulse profile at 1.4\,Ghz from the data used to generate the timing solution.

\subsection{PSR~J0636$+$5129}

The standard filtering described in Section~\ref{sec:obs} leaves 400\ksec\ of exposure for PSR~J0636$+$5129. After folding all events in the 0.25--12.0\keV\ with an ephemeris from NRT observations (Table~\ref{tab:ephem0636}), the H-test significance of 2.5$\sigma$ is not sufficient for a confident claim of detection. However, restricting the energy range to the soft 0.25--2.0\keV\ band brings the detection significance to $4.3\sigma$, and shows a broad single pulse.  Further improvement is obtained with our GTI optimization, which results in $5.5\sigma$ H-test significance for a total of 347\ksec\ of good exposure.  The pulse profile, obtained in the 0.27--0.91\keV\ optimal energy range, is shown in Figure~\ref{fig:profiles}.

\subsection{PSR~J0751$+$1807}

Observations of PSR~J0751$+$1807 were performed with an offset pointing, in an effort to minimize contamination from nearby sources. Using archival \xmm{} data and sources in the 3XMM-DR8 catalog \citep{rosen16}, we determined the exact position to maximize the S/N for the pulsar---see Appendix~\ref{app:pointing0751} for the results of the pointing offset determination, and see \citet{bogdanov19a} for a description of the method.

In the case of PSR~J0751$+$1807, the initial single-trial H-test significance after standard filtering and folding (see ephemeris in Table~\ref{tab:ephem0751}) is $4.1\sigma$ in the full 0.25--12.0\keV\ range. This detection increases when using only events in the 0.25--2.0\keV\ soft band, where thermal emission would be the strongest. There, we obtain a detection significance of 6.5$\sigma$. Finally, as done above for the other pulsars, the GTI optimization further improves the H-test significance: $7.5\sigma$ (0.25--2.0\keV), and $8.5\sigma$ when the optimal energy range of 0.32--1.82\keV\ is used. The pulse profile, resulting from 424\ksec\ of selected GTIs is shown in Figure~\ref{fig:profiles}.  Our confident detection confirms the marginal 1.7$\sigma$ detection claimed from early \xmm{} data of this pulsar \citep{webb04b}.

\subsection{PSR~J1012$+$5307}

A 590\ksec\ exposure of PSR~J1012$+$5307 was obtained with \nicer, after standard filtering.  However, after event folding with the ephemeris of Table~\ref{tab:ephem1012}, the X-ray pulsations remain undetected in the full 0.25--12\keV\ band ($\sim 2\sigma$), and marginally detected ($3\sigma$) when using soft-band photons only.  A fraction of the exposure, early in the mission, was obtained at low Sun angles, close to the instrument's 45$^{\circ}$ limit. Our GTI optimization helps to minimize the impact of exposures with high optical-loading background, such as those at low Sun angles. Applying this optimization improves the H-test significance to $3.7\sigma$ in the 0.25--2.0\keV\ band, and $4.67\sigma$ when also optimizing the energy range---we find that 0.31--1.94\keV\ is optimal.  A total of 548\ksec\ of exposure, out of the 590\ksec\ available, is selected to generate the optimal pulse profile for PSR~J1012$+$5307 (Figure~\ref{fig:profiles}).  Although the detection of the pulsations in the \nicer\ data is somewhat marginal, our observations confirm the $3\sigma$ detection in \xmmlong{} data \citep{webb04b}.  While the apparent X-ray/radio peak separation, $\sim0.3$ phase, is consistent between the \xmm\ and \nicer\ data sets, the double peak structure of the \xmm\ data (Figure 9 of \citealt{webb04b}) is not evident in the \nicer\ data.

\subsection{PSR~J1552$+$5437}

\nicer{} observed this pulsar for a total of 72\ksec\ after standard filtering. Unfortunately, a record-keeping error resulted in the use of incorrect pointing coordinates, offset $\sim2\farcm4$ from the pulsar's reported position \citep{pleunis17}. Therefore, the standard angular distance filtering criterion (excluding $>0\farcm9$) had to be relaxed during the processing; the other standard filtering criteria (Section~\ref{sec:obs}) were retained, resulting in 69\ksec\ of good exposure, but with reduced sensitivity to the off-axis source. After folding with the ephemeris of Table~\ref{tab:ephem1552}, the single-trial detection significance is $\sim1.5\sigma$ whether we choose the full energy range or the soft band only. Applying the GTI optimization increases the significance to $3\sigma$ and $2.8\sigma$ for these two cases, respectively; however, only small fractions of the total exposure (6\ksec\ and 28\ksec, respectively) are retained by the optimization.  Optimizing both sorted GTIs and the energy range also leads to an unrealistically small selection of GTIs (resulting in only 140 events out of $\sim$67000). 

Overall, we conclude that no pulsations were firmly detected from this pulsar. Assuming the detection is real, however, the pulsations would correspond to a pulsed count rate of 0.008\ppm0.004\,c/s (consistent with a non-detection of pulsations), equivalent to a \swift-XRT rate of 0.0004\,c/s.  One would therefore expect 1.2 counts in a 2.9~ks \swift-XRT exposure, which is consistent with the \swift\ non-detection. Longer \nicer\ observations, with the correct pointing, are warranted to confirm or refute the possible presence of X-ray pulsations from this pulsar, or alternatively, longer X-ray imaging observations are required to determine the actual existence of X-ray emission from this pulsar \footnote{\nicer-XTI is a non-imaging instrument, and the detection of a faint pulsar can be confirmed with certainty from the detection of its pulsations.}.

\subsection{PSR~J1744$-$1134}

A total of 71\ksec\ of exposure are available for this pulsar after standard filtering. The ephemeris from the \nrt\ (Table~\ref{tab:ephem1744}) allows us to calculate the pulse phase of each photon. However, whether we choose the full energy band or the 0.25--2.0\keV\ range, or we attempt to optimize it, the detection significances remain $2.3\sigma$, $2.5\sigma$, and $2.7\sigma$, respectively. Optimization of the GTIs to exclude those with high background also results in low-significance detections of pulsations and unrealistically small selections of GTIs ($<$ few ksec).  We therefore conclude that no pulsations are seen from PSR~J1744$-$1134. Past observations of this pulsar, obtained with \chandra, do not have the timing resolution necessary to confirm or refute our conclusions.

\subsection{PSR~J2241$-$5236}
The phase-folding of event times from PSR~J2241$-$5236 leads to pulsations detected with a single-trial significance of $4.5\sigma$ when considering the full \nicer\ band, and $4.9\sigma$ when using 0.25--2.0\keV\ photons.   GTI optimization permits excluding periods of high background and increases the significance to $6.3\sigma$ (0.25--2.0\keV). Finally, the grid search described above to find the optimal energy range yields 0.41--1.14\keV, i.e., excluding the range where the optical loading is the strongest, and resulting in pulsations with a significance at the $7.4\sigma$ confidence level.  Using these 100\ksec\ of optimally selected GTIs, we provide the pulse profile for PSR~J2241$-$5236, along with the \textit{Parkes Radio Telescope} pulse profile, in Figure~\ref{fig:profiles}.

\subsection{PSR~J0740+6620, PSR~J1231$-$1411, and PSR~J1614$-$2230}

The detection of X-ray pulsations from J1231$-$1411 has been reported in \cite{ray19b} and \cite{bogdanov19a}, while PSR~J1614$-$2230 is the subject of another article in preparation \citep{arzoumanian19}, confirming the detection reported in \cite{pancrazi12}. The \nicer{} observations of PSR~J0740+6620 are also presented in \cite{arzoumanian19}.

\begin{deluxetable}{lcccccc}[t]
\tabletypesize{\normalsize}
\tablecaption{Pulsed thermal luminosities, spin-down luminosities and efficiencies\label{tab:edot}}
\tablehead{
\colhead{Pulsar} & \colhead{\nh} & \colhead{$kT_{\rm BB}$\tablenotemark{(a)}}& \colhead{$F^{\rm pulsed}_{0.2-10.0\,{\rm keV}}$} & \colhead{$L^{\rm pulsed}_{0.2-10.0\,{\rm keV}}$} & \colhead{$\dot{E}$} & \colhead{$L_{X}/\dot{E}$} \\
\colhead{} & \colhead{($\ee{20}\percmsq$)} & \colhead{(\keV)}& \colhead{(\cgsflux)} & \colhead{(\cgslum)} & \colhead{(\cgslum)} & \colhead{($\ee{-4}$)} 
}
\startdata
PSR~J0614$-$3329  & 3.0 & 0.23  & 3.5\tee{-14} & 3.0\tee{31} & 2.4\tee{34} & 13.0\\
PSR~J0636$+$5129  & 9.5 & 0.18  & 2.1\tee{-14} & 1.0\tee{29} & 5.6\tee{33} & 0.2\\
PSR~J0751$+$1807  & 6.2 & 0.10 & 3.2\tee{-14} & 4.4\tee{30} & 5.7\tee{33} & 7.7\\
PSR~J1012$+$5307  & 0.8 & 0.26  & 7.8\tee{-15} & 7.7\tee{29} & 2.7\tee{33} & 2.9\\
PSR~J2241$-$5236  & 1.1 & 0.26  & 3.2\tee{-14} & 3.5\tee{30} & 2.5\tee{34} & 1.4\\
\enddata
\tablenotetext{(a)}{ \ \ We assumed a 0.1\keV\ blackbody for PSR~J0751$+$1807, and the values presented in Section~\ref{sec:msp} for the others.}
\tablecomments{The pulsed X-ray fluxes and luminosities reported are unabsorbed. The luminosities are calculated with the distances reported in Table~\ref{tab:targets}.  The $\dot{E}$ values reported have\textbf{} been corrected for the Shklovskii effect \citep{shklovskii70}.}
\end{deluxetable}

\section{Discussion and Conclusion}
\label{sec:discussion}

\subsection{Thermal luminosity vs spin-down luminosity}

From the pulsed count rates listed in Table~\ref{tab:res}, deduced from the pulse profile after GTI and energy range optimizations, we estimated with WebPIMMS\footnote{Available at \url{https://heasarc.gsfc.nasa.gov/cgi-bin/Tools/w3pimms/w3pimms.pl}} the pulsed X-ray unabsorbed flux and luminosity (with distances from Table~\ref{tab:targets}) for the MSPs with detected pulsations. To do so, we used the blackbody temperatures published in the literature (see Section~\ref{sec:msp}, or assuming $kT_{\rm BB}=0.1\keV$ for PSR~J0751$+$1807). We also assumed the value of hydrogen column density \nh\ obtained from neutral H maps using the HEASARC \nh\ tool\footnote{Available at \url{https://heasarc.gsfc.nasa.gov/cgi-bin/Tools/w3nh/w3nh.pl}; Although these values are integrated along the line of sight through the Galaxy, all five pulsars are $>18\deg$ above the Galactic Plane and these \nh\ values should be reasonable estimates.} \citep[with the most recent maps from][]{hi4pi16}.  These pulsed luminosities are listed in Table~\ref{tab:edot} together with spin-down luminosity $\dot{E}$ (corrected for the Shklovskii effect).  We find efficiencies $L_{X}/\dot{E}$ in the range \ee{-5}--\ee{-3}, similar to those of other MSPs \citep[e.g., Figures 9 and 10 in ][]{lee18}, but the values that we report in Table~\ref{tab:edot} should be viewed as lower limits since they were derived from the pulsed flux only.


\subsection{Radio--X-ray phase separation \label{sec:phase}}

Producing properly phase-aligned radio and X-ray pulse profiles can be challenging. For most radio pulsar timing studies an overall absolute phase error is irrelevant, but for alignment with other wavebands, correct absolute timing is critical, so we provide some detail here. In this work, the phase 0.0 is defined by the radio template used to extract the radio TOAs.  The NRT templates are aligned such that the peak of the main pulse is the fiducial point.  Pulsar timing codes (\texttt{TEMPO}, \textsc{Tempo2}, and PINT) produce a fictitious TOA that has zero residual to the model (with its time, observing site, and frequency defined by the \texttt{TZRMJD}, \texttt{TZRSITE}, and \texttt{TZRFRQ} parameters). The PINT \texttt{photonphase} task or the \textsc{Tempo2} \texttt{photons} plugin will assign pulse phases to the X-ray photons using this reference. Possible sources of error include: (1) uncertainty in extrapolating from the radio arrival time to infinite frequency because of dispersion measure uncertainty; (2) uncalibrated cable or pipeline delays in data taking systems; (3) inconsistent application of observatory clock corrections (e.g., confusion over whether \texttt{TZRMJD} is in observatory clock time or UTC); (4) using different ephemerides or positions when barycentering the two datasets, etc.

The phase alignments shown in Figure \ref{fig:profiles} are made with the timing models presented in this paper (Tables~\ref{tab:ephem0614}--\ref{tab:ephem2241}). To provide a cross check for potential errors in the alignment procedure, we reproduced Figure~\ref{fig:profiles} for PSR~J0636$+$5129 and PSR~J1012$+$5307 using timing models from NANOGrav \citep{arzoumanian18}, which are from the Green Bank Telescope, instead of NRT. We obtained very similar profiles and peak separations, giving us confidence that the alignments presented in Figure~\ref{fig:profiles} are correct; in particular, that the clock corrections from NRT are properly taken into account by the PINT {\tt photonphase} task.  

We use the Fourier decompositions presented in Figure~\ref{fig:harmonics} to determine the phase of the X-ray pulses and therefore the phase offset $\Delta\phi$ with the radio peak:
\begin{itemize}
    \item PSR~J0614$-$3329: Both X-ray and radio main peak are broad and essentially aligned ($\Delta\phi\sim0.07$). Furthermore, the secondary radio peaks appear to be somewhat aligned with the secondary X-ray pulse (see Figure~\ref{fig:profiles}).
    \item PSR~J0636$+$5129: The seemingly broad and asymmetric X-ray profile (fast rise and slower decay) of this pulsar leads the radio pulse by $\Delta\phi\sim0.3$--0.4.
    \item PSR~J0751$+$1807: The main radio pulse lags the X-ray peak by $\Delta\phi\sim0.1$. However, a ``precursor'' radio pulse precedes the main radio pulse by $\sim$0.1 in phase, and is therefore aligned with the X-ray pulse.
    \item PSR~J1012$+$5307: This pulsar also displays a large offset, with the X-ray pulse leading the main radio pulse by $\Delta\phi\sim0.3$, while the secondary radio pulses fall in the trough of the X-ray profile.
    \item PSR~J2241$-$5236: The X-ray pulse is broad, with a duty cycle of more than 60\%, and a peak that lags the narrow radio pulse (duty cycle of a few percent at most) by $\Delta\phi\sim0.2$.
\end{itemize}

The $\Delta\phi$ offsets in the range 0.2--0.3 for three out of the five MSPs in our sample are in contrast with other MSPs studied with \nicer. As listed in Section~\ref{sec:intro}, three of the four pulsars presented in \cite{bogdanov19a} all have near perfect alignment between the X-ray and radio pulses (PSR~J0437$-$4715, PSR~J0030$+$0451, and PSR~J1231$-$1411)), while others have  misaligned X-ray and radio pulses, such as PSR~J1614$-$2230.  The large $\Delta\phi\sim 0.2-0.3$ phase offsets between the radio and X-ray measured in this work may put into question the surface heating patterns and the centered dipolar nature of these MSP magnetic fields. For example, the location of the hot regions, separated by $\sim65\deg$ \citep{riley19}, at the surface of PSR~J0030$+$0451 suggest a magnetic field that significantly deviates from a simple centered dipolar field \citep{bilous19}, even though the X-ray and radio pulses are aligned.  Additionally, the observed radio profiles could find their origin in the outer edge (i.e., non-axial) parts of the beam, instead of the core of the beam emission \citep{lyne88,kramer99}, therefore resulting in an offset with the X-ray pulses coming from the footprints of the magnetic field at the NS surface. Overall, the relationship between the X-ray and radio pulsations of thermally-emitting MSPs therefore ought to be studied in more detail, and a more systematic study of radio and X-ray profiles could help constrain the complexity of the magnetic fields in these sources. The phase alignment of the radio and X-ray pulses with the $\gamma$-ray pulsations may also be informative, although the latter originates from emission further out in the magnetosphere and does not necessarily align with the radio or X-ray profiles.

\subsection{Harmonics in the pulse profiles}
\label{sec:harm}

\begin{figure*}[t]
\begin{center}
\includegraphics[width=0.5\textwidth]{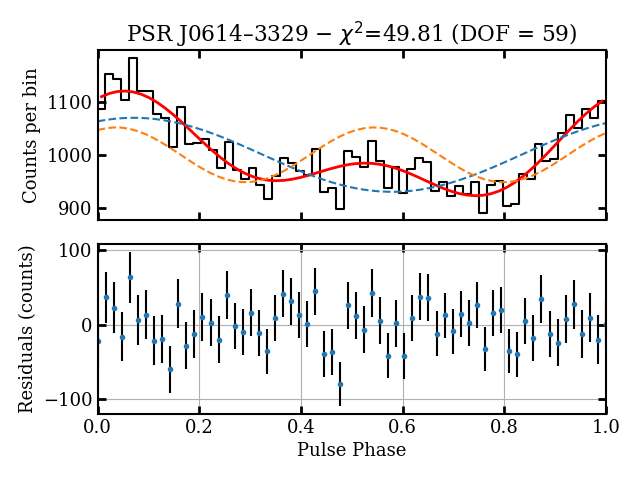}~
\includegraphics[width=0.5\textwidth]{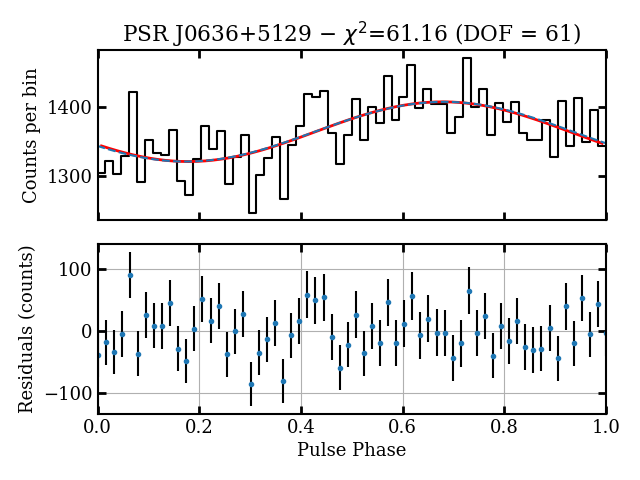}
\includegraphics[width=0.5\textwidth]{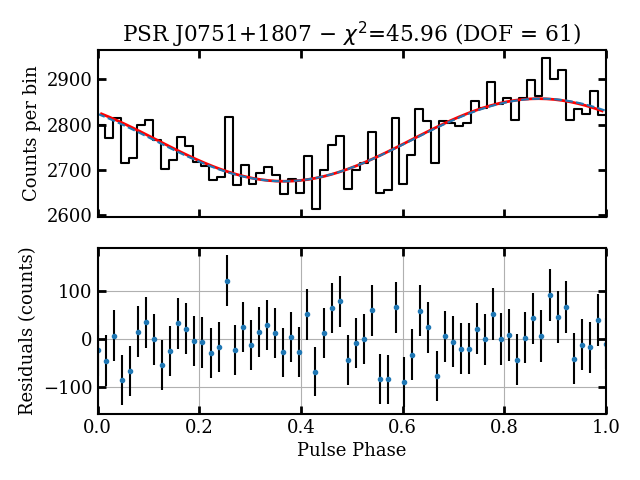}~
\includegraphics[width=0.5\textwidth]{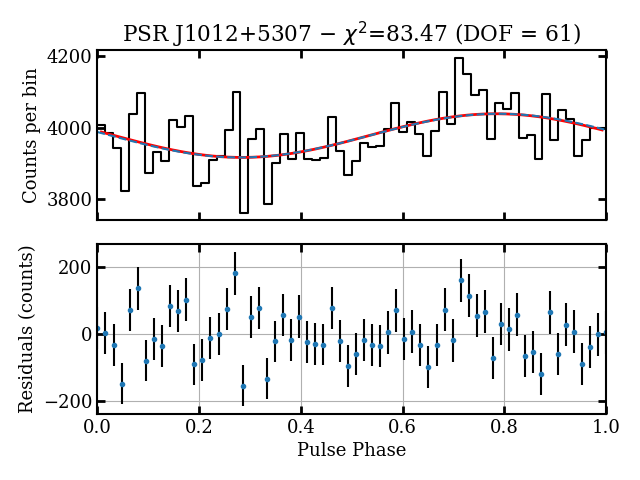}
\includegraphics[width=0.5\textwidth]{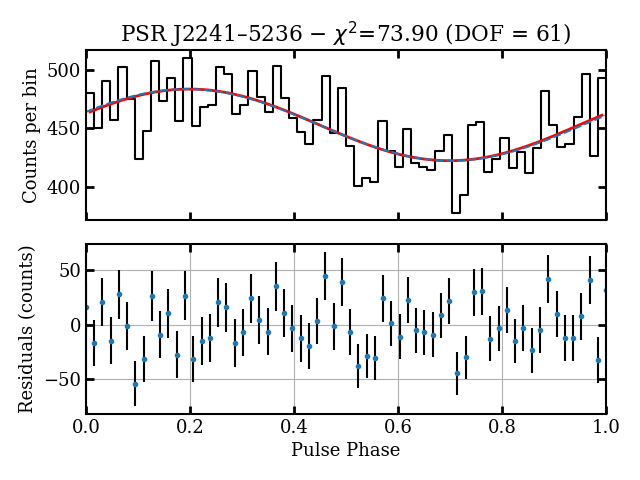}
\caption{Fourier decomposition of the \textit{NICER} pulse profiles of the five MSPs with detected pulsations, with 64 phase bins. Only PSR~J0614$-$3329 required more than one harmonic. In each panel, the dashed lines are the harmonic components, and the red solid line is the sum of the components. The bottom plot of each panel shows the residuals. Only one rotational cycle is shown.
}
\label{fig:harmonics}
\end{center}
\end{figure*}

For the five pulsars with detected pulsations, we apply a Fourier decomposition technique to search for the presence of harmonics in the broad, sine-like profiles.  We find that PSR~J0614$-$3329 is the only one that requires two harmonics (using a single harmonic leaves structured residuals and gives $\chisqnu > 2$). For the other four pulsars, a single sine function is sufficient to describe the pulse profile (i.e., $\chisqnu \approx 1$ and white residuals). Figure~\ref{fig:harmonics} shows the harmonic fits and residuals for the five pulsars. 

The detection of higher harmonics may provide evidence for the various effects that distort the pulses from a simple sinusoid (See below) and/or for the presence of a second visible polar cap (i.e., a secondary pulse, as clearly seen in the case of PSR~J0030$+$0451, \citealt{bogdanov09}).  At spin frequencies up to $\sim300$~Hz, atmosphere beaming (since MSP surface emission is not isotropic), as well as self-occultation of the polar caps by the star, make the pulse profile deviate from a sinusoidal shape. Above 300~Hz, other effects, such as Doppler boosting and aberration, add more power to the harmonic content \citep{miller15}.

The two harmonics in PSR~J0614$-$3329 can be readily interpreted as due to the presence of a secondary pulse, about half a phase from the dominant pulse (see Figures~\ref{fig:profiles} and \ref{fig:harmonics}). This likely originates from a second polar cap, similar to that observed in PSR~J1231$-$1411 and PSR~J0030$+$0451  \citep{bogdanov19a}. For the other four pulsars of the present study, the single harmonic detected may simply be ascribed to the low S/N of the data sets.  For comparison, the key \nicer\ targets used for pulse profile modeling (PSR~J0437$-$4715, PSR~J0030$+$0451, PSR~J1231$-$1411, and PSR~J2124$-$3358) show up to four harmonics in their pulse profiles \citep{bogdanov19a}, with high S/N provided by deep ($\sim$1--2\,Ms) exposures.  Observations with exposures of 1--5\,Ms for the five pulsars reported here will likely reveal their full harmonic content and is required to perform pulse profile modeling analyses to extract mass and radius constraints.

\subsection{Conclusion and perspectives}

We present a systematic search for X-ray pulsations using \nicer\ timing observations of a set of seven nearby rotation-powered MSPs. X-ray pulsations are securely detected from five of the targets.  Specifically, the single-trial significance is $>4.3\sigma$ for all MSPs except for PSR~J1012$+$5307 ($3\sigma$). This significance is improved ($\geq4.7\sigma$ in all cases) when optimizing the selection of GTIs to favor low background count rates (see Section~\ref{sec:gtisort}). Although the number of trials involved with the GTI sorting method is not straightforward to quantify exactly, it is much less than the number of GTIs for a given MSP data set. Indeed, since they are sorted by increasing total rate, they are highly correlated, making the trial factor for our GTI sorting method on the order of a few.

Because of the low count rates for the MSPs in the present paper (see Table~\ref{tab:res}), a proper spectral analysis was unfortunately not possible. Using a \nicer{} background model \citep{ray19b,bogdanov19a}, we estimate that $\lesssim$~5--10\% of the total detected counts in the soft X-ray band (0.5--2.0\keV) originate from the observed pulsars. Given the uncertainties involved with modeling the background\footnote{\nicer\ is a non-imaging instrument and the background is modeled based on observations of blank-sky exposures, as well as house-keeping and space-weather parameters.}, we refrain from providing spectral parameters for these faint pulsars. Imaging instruments sensitive in the soft X-ray band, such as \xmmlong{}, \textit{eROSITA} \citep{predehl14}, or the future \textit{Athena X-ray Observatory} \citep{nandra13} are better suited for such spectral analyses. 

The results of this paper nonetheless enlarge the sample of radio MSPs detected as pulsed X-ray sources---in particular, those with thermal emission. All five MSPs detected have broad pulsed profiles, fitted with one or two harmonics (see Section~\ref{sec:harm}).  Published studies of these MSPs showed that they also have X-ray spectra consistent with thermal emission\footnote{In all five cases, the spectral analyses, reported in the literature, were performed on rather low S/N data from \swift-XRT, \xmm, or \chandra, with less than a few hundred of counts.} (simple or double black body, see Section~\ref{sec:msp} for the details).  The broad pulses and the spectral shapes lead us to tentatively conclude that we observe thermal X-ray pulsations from these five MSPs.  We note that more detailed spectral analyses, with high S/N, is necessary to confirm with certainty the spectral nature of these pulsations.  Indeed, PSR~J0218$+$4232 displays a pair of rather broad X-ray pulses ($\sim0.25$ phase), connected by a "bridge" joining them \citep{deneva19}. However, its X-ray spectrum is hard and purely non-thermal (with a power-law photon index $\Gamma\approx1.1$), making the detection of pulsations possible up to $\sim10\keV$ in the \nicer\ data \citep{webb04a,rowan19}.  To confirm the thermal nature of the MSPs studied here would require observations with spectro-imaging X-ray instruments.

The detection of pulsations, if they are indeed of thermal origin, seems to confirm that for the majority of MSPs the main source of their X-ray emission comes from their heated polar caps.  The findings of this Letter also make these MSPs promising targets for follow-up X-ray observations with \nicer\ and/or with future proposed telescopes such as STROBE-X \citep{ray19a} or \textit{eXTP} \citep{watts19}, to enable constraints on the NS mass–radius relation and the dense matter EOS via the pulse profile modeling technique. The binary MSPs with independent mass measurements from radio timing, PSR~J0751$+$1807 and PSR~J1012$+$5307, are of particular interest in this sense since they can offer more stringent constraints on the NS radius.  PSR~J0614$-$3329, the brightest of the five MSPs studied here, is also a promising target.

Following the discoveries of X-ray pulsations from five MSPs, \nicer\ continues to observe these targets to better characterize their X-ray pulse profiles.  These MSPs will require a few Ms of exposure per pulsar with \nicer\. Furthermore, \nicer\ will also target newly discovered MSPs such as those discovered by the Arecibo PALFA survey \citep{parent19}.

\acknowledgments

The authors are grateful to the referee for their insightful suggestions that contributed to improving this paper. S.G. acknowledges the support of the French Centre National d'\'{E}tudes Spatiales (CNES). C. M. is a NASA Postdoctoral fellow. This work was supported in part by NASA through the \textit{NICER} mission and the Astrophysics Explorers Program. This research has made use of data and software provided by the High Energy Astrophysics Science Archive Research  Center (HEASARC), which is a service of the Astrophysics Science Division at NASA/GSFC and the High Energy Astrophysics Division of the Smithsonian Astrophysical Observatory.  We acknowledge extensive use of NASA's Astrophysics Data System (ADS) Bibliographic Services and the ArXiv. The Nan\c{c}ay Radio Observatory is operated by the Paris Observatory, associated with the French Centre National de la Recherche Scientifique (CNRS). The National Radio Astronomy Observatory is a facility of the National Science Foundation operated under cooperative agreement by Associated Universities, Inc. S.M.R is a CIFAR Fellow and is supported by the NSF Physics Frontiers Center award 1430284.  We acknowledge financial support from the {\it Programme National de Cosmologie et Galaxies} (PNCG), {\it Programme National Hautes Energies} (PNHE), and {\it Programme National Gravitation, R\'{e}f\'{e}rences, Astronomie, M\'{e}trologie} (PNGRAM) of CNRS/INSU, France. This work benefited from the support of the {\it Entretiens sur les Pulsars} funded by Programme National High Energies (PNHE) of CNRS/INSU with INP and IN2P3, co-funded by CEA and CNES.

\vspace{5mm}

\facilities{\nicer, \nrt, \textit{Parkes Radio Telescope}}

\software{
\texttt{astropy} (\citealt{astropy18}, \url{https://ascl.net/1304.002}), \texttt{PINT} (\citealt{luo19}, \url{https://ascl.net/1902.007}), \texttt{TEMPO2} (\citealt{hobbs12}, \url{https://ascl.net/1210.015}), \texttt{HEAsoft} (\citealt{heasoft14}, \url{https://ascl.net/1408.004}), \texttt{NICERsoft}, (\url{https://github.com/paulray/NICERsoft})
}

\bibliographystyle{aasjournal}
\bibliography{biblio}


\appendix

\section{Pointing optimization for PSR~J0751$+$1807}
\label{app:pointing0751}

\begin{figure*}[t]
\begin{center}
\includegraphics[width=0.42\textwidth]{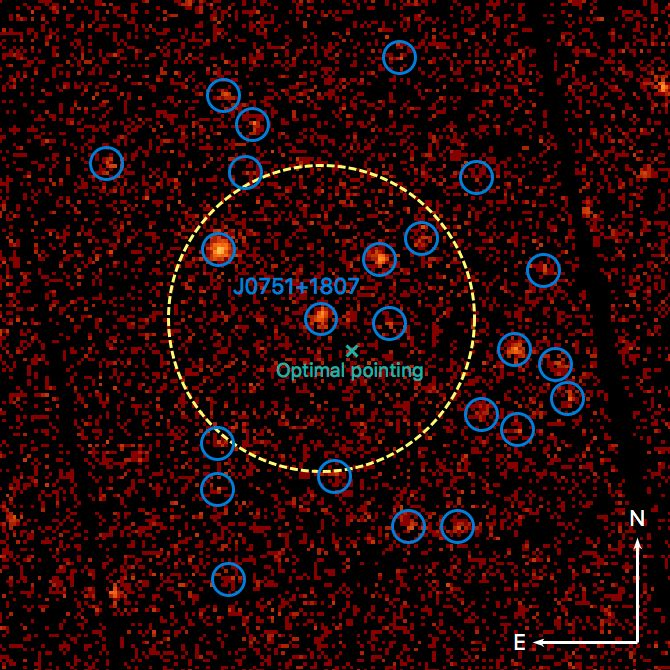}
\includegraphics[width=0.57\textwidth]{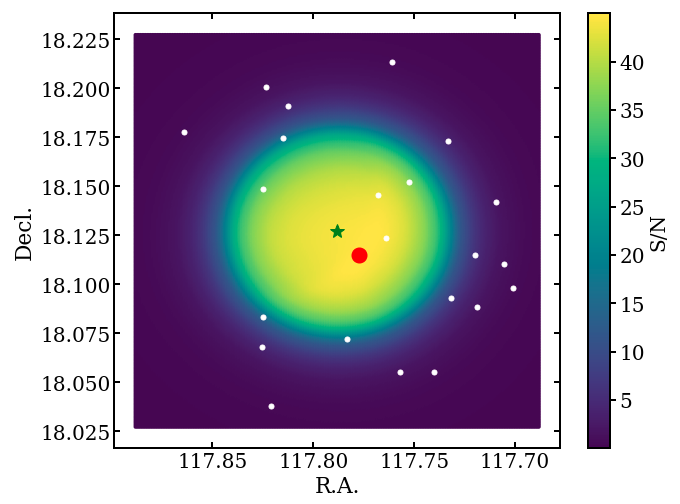}
\caption{({\it left}) \textit{XMM-Newton} EPIC MOS image of PSR~J0751$+$1807 and nearby sources. The teal '$\times$' shows the position of the optimal \textit{NICER} pointing that maximizes the S/N from the pulsar, i.e., minimizes the contamination from all other sources (blue circles) within 6\arcmin. The dashed yellow circle shows the size of the \nicer{} 6\farcm2 half-power diameter point spread function.  ({\it right}) Map of the S/N of PSR~J0751$+$1807 as a function of \textit{NICER} pointing.  The green star shows the pulsar position, and the red circle shows the calculated optimal pointing position that maximizes the S/N. The optimal pointing is 1\arcmin{} from the pulsar position and permits a gain in S/N of a few percent.}
\label{fig:J0751pointing}
\end{center}
\end{figure*}

Figure~\ref{fig:J0751pointing} in this appendix presents the results of the pointing optimization for PSR~J0751$+$1807, as described in \citep{bogdanov19a}.



\section{Millisecond pulsar ephemerides}
\label{app:ephem}

In the following tables (\ref{tab:ephem0614}--\ref{tab:ephem2241}), we present the ephemerides from the NRT, the \textit{Parkes Radio Telescope}, and \textit{LOFAR} used to calculated the phases of X-ray events detected by \nicer. We stress that these ephemerides were obtained for the sole purpose of phase-folding X-ray events, and should \emph{not} be used as long-term timing solutions for these pulsars.

\begin{deluxetable}{lr}[h]
\tabletypesize{\small}
\tablecaption{Ephemeris of PSR~J0614$-$3329 used in this work and obtained from observations with the \nrt.  The digit in parenthesis represents the $1\sigma$ uncertainty on the last digit. \label{tab:ephem0614}}
\tablehead{ \colhead{Parameter} & \colhead{Value} }
\startdata
Pulsar name\dotfill						                                    & J0614$-$3329 \\
Right Ascension (J2000) \dotfill 			                            	& 06:14:10.347818(8) \\ 
Declination (J2000) \dotfill 					                            & $-$33:29:54.1161(1) \\ 
Proper motion in R.A. (mas yr$^{-1}$) \dotfill 		                    	& 0.61(3) \\ 
Proper motion in Decl. (mas yr$^{-1}$) \dotfill			                    & $-$1.74(4) \\ 
Epoch of position \dotfill   					                            & 56000.0 \\
\hline
Spin frequency, $\nu$ (Hz) \dotfill				                            & 317.594455162825(3) \\ 
Spin frequency derivative, $\dot{\nu}$ (s$^{-2}$)\dotfill 	                & $-1.75678(1)\tee{-15}$ \\ 
Epoch of period  \dotfill 					                                & 56000.0\\
Dispersion measure, DM (pc\,cm$^{-3}$) \dotfill 		                    & 37.050(5) \\ 
Dispersion measure derivative, DM1 \dotfill 		                    	& 0.0004(1) \\ 
Epoch of DM \dotfill   						                                & 56000.0 \\
\hline
Binary model \dotfill 						                                & BT \citep{blandford76} \\
Binary orbital period, $P_{\rm{b}}$ (days)\dotfill 	                    	& 53.584612528(3) \\ 
Projected semi-major axis of orbit, $x$ (lt-s)\dotfill 	                	& 27.6387906(2) \\ 
Orbital eccentricity, $e$\dotfill 			                            	& 0.00018080(1) \\ 
Epoch of periastron, T0 (MJD) \dotfill 				                        & 54986.0699(6) \\ 
Longitude of periastron, $\omega$ (degrees)\dotfill 		                & 15.938(4) \\ 
\hline
Terrestrial time standard (CLK)  \dotfill      			                    & TT(BIPM2011) \\
Reference epoch (MJD)  \dotfill	    	 	  	                        	& 57236.403056165037679 \\
Frequency of reference TOA (MHz) \dotfill		                        	& 1299.619019 \\ 
Solar System ephemeris  \dotfill						                                    & DE436 \\
\enddata
\end{deluxetable}

\begin{deluxetable}{lr}[h]
\tablecaption{Ephemeris of PSR~J0636$+$5129 used in this work and obtained from observations with the \nrt. Digits in parentheses represent the $1\sigma$ uncertainty on the last quoted digit of a parameter value.  \label{tab:ephem0636}}
\tablehead{ \colhead{Parameter} & \colhead{Value} }
\startdata
Pulsar name\dotfill						                                    & J0636$+$5129 \\
Right Ascension (J2000) \dotfill 			                            	& 06:36:04.84618(1) \\ 
Declination (J2000) \dotfill 					                        	& +51:28:59.9651(3) \\ 
Proper motion in R.A. (mas yr$^{-1}$) \dotfill 			                	& 3.22(3) \\ 
Proper motion in Decl. (mas yr$^{-1}$) \dotfill			                	& $-$1.61(6) \\ 
Epoch of position \dotfill   					                        	& 56000.0 \\
\hline
Spin frequency, $\nu$ (Hz) \dotfill				                        	& 348.559231746222(3) \\ 
Spin frequency derivative, $\dot{\nu}$ (s$^{-2}$)\dotfill 	                & $-4.1902(2)\tee{-16}$ \\ 
Epoch of period  \dotfill 				                        		    & 56000.0\\
Dispersion measure, DM (pc\,cm$^{-3}$) \dotfill 		                	& 11.0982(1) \\ 
Epoch of DM \dotfill   					                            		& 56000.0 \\
\hline
Binary model \dotfill 					                	            	& ELL1 \\
Binary orbital period, $P_{\rm{b}}$ (days)\dotfill 		                	& 0.06655133843(7) \\ 
Binary orbital period derivative, $\dot{P}_{\rm{b}}$ \dotfill               & $1.89(5)\tee{-12}$ \\ 
Projected semi-major axis of orbit, $x$ (lt-s)\dotfill 		                & 0.00898621(8) \\ 
Epoch of ascending node passage, T$_{\rm asc}$ (MJD) \dotfill             	& 56027.2483387(8) \\ 
First Laplace parameter, $\sin\,\omega$ \dotfill   		                	& $0.25(18)\tee{-4}$ \\ 
Second Laplace parameter $\cos\,\omega$ \dotfill   		                	& $0.32(18)\tee{-4}$ \\ 
\hline
Terrestrial time standard (CLK)  \dotfill                              		& TT(BIPM2011) \\
Reference epoch (MJD)  \dotfill	    	 	                          		& 57551.55910304646476 \\
Frequency of reference TOA (MHz) \dotfill	                        		& 1547.8940430 \\
Solar System Ephemeris  \dotfill					                        	        	& DE436 \\
\enddata
\end{deluxetable}

\begin{deluxetable}{lr}[h]
\tablecaption{Ephemeris of PSR~J0751$+$1807 used in this work and obtained from observations with the \nrt.  Digits in parentheses represent the $1\sigma$ uncertainty on the last quoted digit of a parameter value. \label{tab:ephem0751}}
\tablehead{ \colhead{Parameter} & \colhead{Value} }
\startdata
Pulsar name\dotfill						                                    & J0751$+$1807 \\
Right Ascension (J2000) \dotfill 				                            & 07:51:09.15481(1) \\ 
Declination (J2000) \dotfill 					                            & +18:07:38.4485(8) \\ 
Proper motion in R.A. (mas yr$^{-1}$) \dotfill 			                    & $-$2.65(4) \\ 
Proper motion in Decl. (mas yr$^{-1}$) \dotfill			                    & $-$13.0(2) \\ 
Epoch of position \dotfill   					                            & 56000.0 \\
\hline
Spin frequency, $\nu$ (Hz) \dotfill				                            & 287.457858396603(3) \\ 
Spin frequency derivative, $\dot{\nu}$ (s$^{-2}$)\dotfill 	                & $6.4353(3)\tee{-16}$ \\ 
Epoch of period  \dotfill 						                            & 56000.0\\
Dispersion measure, DM (pc\,cm$^{-3}$) \dotfill 		                    & 30.2437(2) \\ 
Dispersion measure derivative, DM1 \dotfill 			                    & 0.0007(2) \\ 
Dispersion measure second derivative, DM2 \dotfill 		                    & $-$0.00019(3) \\ 
Epoch of DM \dotfill   							                            & 56000.0 \\
\hline
Binary model \dotfill 							                            & ELL1 \\
Binary orbital period, $P_{\rm{b}}$ (days)\dotfill 		                    & 0.26314426668(3) \\ 
Binary orbital period derivative, $\dot{P}_{\rm{b}}$ \dotfill               & $-2.8(6)\tee{-14}$ \\ 
Projected semi-major axis of orbit, $x$ (lt-s)\dotfill 		                & 0.3966139(4) \\ 
Rate of change of projected semi-major axis, $\dot{x}$ \dotfill             & $-7(8)\tee{-16}$ \\ 
Epoch of ascending node passage, T$_{\rm asc}$ (MJD) \dotfill 	            & 51800.2157344(4) \\ 
First Laplace parameter, $\sin\,\omega$ \dotfill   		                    & $2.7(3)\tee{-6}$ \\ 
Second Laplace parameter $\cos\,\omega$ \dotfill   		                    & $0.3(3)\tee{-6}$ \\ 
\hline
Terrestrial time standard (CLK)  \dotfill      			                    & TT(BIPM2011) \\
Reference epoch (MJD)  \dotfill	    	 	  			                    & 57214.533388622116 \\
Frequency of reference TOA (MHz) \dotfill				                    & 1704.5 \\
Solar System ephemeris  \dotfill						    	                            & DE436 \\
\enddata
\end{deluxetable}

\begin{deluxetable}{lr}[h]
\tablecaption{Ephemeris of PSR~J1012$+$5307 used in this work and obtained from observations with the \nrt.  Digits in parentheses represent the $1\sigma$ uncertainty on the last quoted digit of a parameter value. \label{tab:ephem1012}}
\tablehead{ \colhead{Parameter} & \colhead{Value} }
\startdata
Pulsar name\dotfill						    	                            & J1012$+$5307 \\
Right Ascension (J2000) \dotfill 				                            & 10:12:33.438318(4) \\ 
Declination (J2000) \dotfill 					                            & +53:07:02.23033(4) \\
Proper motion in R.A. (mas yr$^{-1}$) \dotfill 			                    & 2.665(9) \\ 
Proper motion in Decl. (mas yr$^{-1}$) \dotfill			                    & $-$25.50(1) \\ 
Parallax (mas) \dotfill     	       					                    & 1.0(1) \\ 
Epoch of position \dotfill   					                            & 56000.0 \\
\hline
Spin frequency, $\nu$ (Hz) \dotfill				                            & 190.2678373381230(3) \\ 
Spin frequency derivative, $\dot{\nu}$ (s$^{-2}$)\dotfill 	                & $-6.20036(3)\tee{-16}$  \\ 
Epoch of period  \dotfill 						                            & 56000.0\\
Dispersion measure, DM (pc\,cm$^{-3}$) \dotfill 		                    & 9.02169(9) \\ 
Epoch of DM \dotfill   							                            & 56000.0 \\
\hline
Binary model \dotfill 							                            & ELL1 \\
Binary orbital period, $P_{\rm{b}}$ (days)\dotfill 		                    & 0.604672713901(5) \\ 
Projected semi-major axis of orbit, $x$ (lt-s)\dotfill 		                & 0.58181784(4) \\ 
Epoch of ascending node passage, T$_{\rm asc}$ (MJD) \dotfill 	            & 50700.08162698(5) \\ 
First Laplace parameter, $\sin\,\omega$ \dotfill   		                    & $1.2(1)\tee{-6}$ \\ 
Second Laplace parameter $\cos\,\omega$ \dotfill   		                    & $0.2(1)\tee{-6}$ \\ 
\hline
Terrestrial time standard (CLK)  \dotfill      			                    & TT(BIPM2011) \\
Reference epoch (MJD)  \dotfill	    	 	  			                    & 57211.626417018760524 \\
Frequency of reference TOA (MHz) \dotfill				                    & 1419.782959 \\ 
Solar System ephemeris  \dotfill							                                & DE436 \\
\enddata
\end{deluxetable}

\begin{deluxetable}{lr}
\tablecaption{Ephemeris of PSR~J1552$+$5437 used in this work and obtained from observations with \textit{LOFAR} and \fermi\ (from \citealt{pleunis17}). Digits in parentheses represent the $1\sigma$ uncertainty on the last quoted digit of a parameter value.  \label{tab:ephem1552}}
\tablehead{ \colhead{Parameter} & \colhead{Value} }
\startdata
Pulsar name\dotfill							                                & J1552$+$5437 \\
Right Ascension (J2000) \dotfill 				                            & 15:52:53.3311(2) \\ 
Declination (J2000) \dotfill 					                            & +54:37:05.787(1) \\ 
Epoch of position \dotfill   					                            & 56285 \\
\hline
Spin frequency, $\nu$ (Hz) \dotfill				                            & 411.8805314243(1) \\
Spin frequency derivative, $\dot{\nu}$ (s$^{-2}$)\dotfill 	                & $-4.74(2)\tee{-16}$ \\
Epoch of period  \dotfill 					                                & 56285 \\
Dispersion measure, DM (pc\,cm$^{-3}$) \dotfill 		                    & 22.9000(5) \\
Epoch of DM \dotfill   						                                & 56285 \\
\hline
Terrestrial time standard (CLK)  \dotfill      			                    & TT(BIPM2011) \\
Solar System ephemeris  \dotfill						                                    & DE421 \\
\enddata
\end{deluxetable}

\begin{deluxetable}{lr}[h]
\tablecaption{Ephemeris of PSR~J1744$-$1134 used in this work and obtained from observations with the \nrt.  Digits in parentheses represent the $1\sigma$ uncertainty on the last quoted digit of a parameter value. \label{tab:ephem1744}}
\tablehead{ \colhead{Parameter} & \colhead{Value} }
\startdata
Pulsar name\dotfill						                                    & J1744$-$1134 \\
Right Ascension (J2000) \dotfill 				                            & 17:44:29.411066(1) \\
Declination (J2000) \dotfill 					                            & $-$11:34:54.72002(9) \\
Proper motion in R.A. (mas yr$^{-1}$) \dotfill 			                    & 18.783(6) \\ 
Proper motion in Decl. (mas yr$^{-1}$) \dotfill			                    & $-$9.28(3) \\ 
Parallax (mas) \dotfill     	       					                    & 2.78(5) \\ 
Epoch of position \dotfill   					                            & 56000.0 \\
\hline
Spin frequency, $\nu$ (Hz) \dotfill				                            & 245.4261234486887(2) \\ 
Spin frequency derivative, $\dot{\nu}$ (s$^{-2}$)\dotfill 	                & $-5.38094(2)\tee{-16}$ \\ 
Epoch of period  \dotfill 						                            & 56000.0\\
Dispersion measure, DM (pc\,cm$^{-3}$) \dotfill 		                    & 3.13845(6) \\ 
Epoch of DM \dotfill   							                            & 56000.0 \\
\hline
Terrestrial time standard (CLK)  \dotfill      			                    & TT(BIPM2011) \\
Reference epoch (MJD)  \dotfill	    	 	  			                    & 57210.939926879214401 \\
Frequency of reference TOA (MHz) \dotfill				                    & 1681.251953 \\ 
Solar System ephemeris  \dotfill							                                & DE436 \\
\enddata
\end{deluxetable}

\begin{deluxetable}{lr}[h]
\tablecaption{Ephemeris of PSR~J2241$-$5236 used in this work and obtained from observations with the Parkes Radio Telescope.  Digits in parentheses represent the $1\sigma$ uncertainty on the last quoted digit of a parameter value.  \label{tab:ephem2241}}
\tablehead{ \colhead{Parameter} & \colhead{Value} }
\startdata
Pulsar name\dotfill						                                    & J2241$-$5236 \\
Right Ascension (J2000) \dotfill 				                            & 22:41:42.016912(2) \\
Declination (J2000) \dotfill 					                            & $-$52:36:36.21305(1) \\
Proper motion in R.A. (mas yr$^{-1}$) \dotfill 			                    & 18.840(3) \\
Proper motion in Decl. (mas yr$^{-1}$) \dotfill			                    & $-$5.269(3) \\
Parallax (mas)  \dotfill    	       					                    & 0.83(3) \\
Epoch of position \dotfill   					                            & 55044.15587 \\
\hline
Spin frequency, $\nu$ (Hz) \dotfill				                            & 457.3101568473538(3) \\
Spin frequency derivative, $\dot{\nu}$ (s$^{-2}$)\dotfill 	                & $-1.442288(2)\tee{-15}$ \\
Epoch of period  \dotfill 						                            & 55044.15587 \\
Dispersion measure, DM (pc\,cm$^{-3}$) \dotfill 		                    & 11.44(1) \\
Epoch of DM \dotfill   							                            & 55044.15587 \\
\hline
Binary model \dotfill 							                            & ELL1 \\ 
Binary orbital period, $P_{\rm{b}}$ (days)\dotfill 		                    & 0.145672237818(2) \\ 
Time derivative of orbital frequency (Hz s$^{-1}$) \dotfill                 & $-3.17(3)\tee{-21}$ \\
Projected semi-major axis of orbit, $x$ (lt-s)\dotfill 		                & 0.02579536(1) \\
Rate of change of projected semi-major axis \dotfill 		                & $1.3(1)\tee{-15}$ \\
Epoch of ascending node passage, T$_{\rm asc}$ (MJD) \dotfill 	            & 56726.96359375(1) \\
First Laplace parameter, $\sin\,\omega$ \dotfill   		                    & $3(8)\tee{-7}$ \\ 
Second Laplace parameter $\cos\,\omega$ \dotfill   		                    & $7(8)\tee{-7}$ \\
\hline
Terrestrial time standard (CLK)  \dotfill      			                    & TT(TAI) \\
Reference epoch (MJD)  \dotfill	    	 	  			                    & 57831.06107389881430 \\
Solar System ephemeris  \dotfill						                                    & DE421 \\
\enddata
\end{deluxetable}



\end{document}